\documentclass[12pt,preprint]{aastex}

\slugcomment{To be published in the November 20, 2003, Vol 598 issue of the Astrophysical Journal}

\received{2003 March 2}
\revised{2003 July 4}
\begin{document}
\title{GAMMA-RAY SPECTRA \& VARIABILITY OF THE CRAB NEBULA EMISSION OBSERVED BY BATSE}
\author{J. C. Ling }
\affil{Jet Propulsion Laboratory 169-327, California Institute of Technology}
\affil{4800 Oak Grove Drive, Pasadena, CA 91109}
\email{james.c.ling@jpl.nasa.gov}
\and
\author{Wm. A. Wheaton }
\affil{ Infrared Processing and Analysis Center, California Institute 
of Technology }
\affil{100-22, Pasadena, CA 91125}
\email{ waw@ipac.caltech.edu}

\begin{abstract}

We report $\sim$600 days of BATSE earth-occultation observations of the
total gamma-ray (30 keV to 1.7 MeV) emission from the Crab nebula, between
1991 May 24 (TJD 8400) and 1994 October 2 (TJD 9627).  Lightcurves from
\{35--100, 100--200, 200--300, 300--400, 400--700, and 700--1000\} keV,
show that positive fluxes were detected by BATSE in each of these six
energy bands at significances of approximately \{ 31, 20, 9.2, 4.5, 2.6,
and 1.3\} $\sigma$ respectively per day. We also observed significant flux
and spectral variations in the 35--300 keV energy region, with time scales
of days to weeks.  The spectra below 300 keV, averaged over typical CGRO
viewing periods of 6--13 days, can be well described by a broken power law
with average indices of $\sim$2.1 and $\sim$2.4 varying around a spectral
break at $\sim$100 keV.  Above 300 keV, the long-term averaged spectra,
averaged over three 400 d periods (TJD 8400-8800, 8800-9200, and
9200-9628, respectively) are well represented by the same power law with index of $\sim$2.34 up to $\sim$670 keV, plus a hard spectral
component extending from $\sim$670 keV to $\sim$1.7 MeV, with a spectral
index of $\sim$1.75. The latter component could be related to a complex
structure observed by COMPTEL in the 0.7-3 MeV range.  Above 3 MeV, the
extrapolation of the power-law continuum determined by the low-energy
BATSE spectrum is consistent with fluxes measured by COMPTEL in the 3-25
MeV range, and by EGRET from 30-50 MeV.  We interpret these results as
synchrotron emission produced by the interaction of particles ejected from
the pulsar with the field in different dynamical regions of the nebula
system, as observed recently by HST, XMM-Newton, and Chandra.

\end{abstract}
\keywords{gamma rays: observations --- ISM: individual (Crab Nebula) --- supernova remnants}


\section{INTRODUCTION}

Since the birth of gamma-ray astronomy in the early 1960's, the Crab nebula has been one of the premier sources for observations by all gamma-ray experiments, operating in the wide spectral range from tens of keV to TeV, with ground-based (100 GeV to 1 TeV), balloon or satellite-borne instruments (30 keV to 1 GeV). In the soft gamma-ray region between 30 keV and 2 MeV covered by BATSE, the Crab nebula is one of the two brightest sources, along with Cygnus X-1, in the sky (see Ling et al. 2000). Because of the relatively steady emission seen over the years, it has frequently been used as a calibration source for new satellite missions. 

The Crab spectrum of the total emission (nebula and pulsar) in the 30 keV - 2 MeV range has been measured over the last three decades by a series of balloon (Kurfess 1971; Laros, Matteson $\&$ Pelling 1973; Walraven et al. 1975, Gruber $\&$ Ling 1977; Leventhal, MacCallum $\&$ Watts 1977; Ling et al. 1977; Ling et al. 1979; Strickman et al. 1979; Manchanda et al. 1982; Graser $\&$ Schoenfelder 1982; Pelling et al. 1987; McConnell et al. 1987; Ubertini et al. 1994) and satellite experiments (Ariel V: Carpenter, Coe $\&$ Engel 1976; OSO 8: Dolan et al. 1977; HEAO-1: Knight, F. K. 1982, Jung 1989; HEAO-3: Mahoney, Ling $\&$ Jacobson 1984; CGRO: Ulmer et al. 1995; Much et al. 1996, van der Meulen et al. 1998; Kuiper et al. 2001). Below 300 keV, the spectrum was initially thought to be a single power law based on earlier measurements by Walraven et al. (1975), Carpenter, Coe and Engel (1976), Dolan et al. (1977), and Ling et al. (1979) with an index of $\sim$2.1. Laros, Matteson and Pelling (1973) were the first to observe a change of the spectrum below $\sim$100 keV. This was confirmed later by Strickman et al. (1979) and Hasinger (1984). Strickman et al. (1979) showed that the spectral index changed from 1.99 to $\sim$2.42 at $\sim$80 keV based on a balloon measurement. Jung (1989), using the HEAO-1 data from three separate observations in September 1977, March 1978, and September 1978, showed possibly a variable spectral break at 130 $\pm$ 9, 139 $\pm 7$, and 118 $\pm$ 5 keV, respectively, with the spectral index changed from $\sim$2.08 to $\sim$2.50. Observations of a possible cyclotron feature at 73 - 79 keV in both pulsed and total emission of the Crab spectra were also reported by several groups (Ling et al. 1979; Strickman et al.1982; Manchanda et al., 1982; Ayre et al. 1983; Watanabe 1985). Because this feature was not consistently observed by other experiments (see Owen 1991 for a review), it is considered a variable feature until a more definitive measurement by INTEGRAL can be obtained in the near future. 

Between 300 keV and 2 MeV, the spectrum is not well established due to reports of several features/structures in the literature over the past three decades. These include: 

(1) A variable red-shifted annihilation line centered at energy between 400 and 450 keV superposed on the power-law continuum. Positive detections for such a feature were reported by Leventhal, MacCallum $\&$ Watts, 1977 (1977), Yoshimori et al.(1979), Jung (1989), Ayre et al.(1983), Agrinier et al.(1990), Parlier et al.(1991), and no detections were reported by Ling et al.(1977), Mahoney, Ling $\&$ Jacobson (1984),  Hameury et al. (1983), and McConnell et al. (1987) (see Owen 1991 for a review) 

(2) A possible blue shifted annihilation feature centered at energy $\sim$540 keV was also reported by Granat-SIGMA (Gilfanov et al, 1994). However, this was not confirmed by BATSE's simultaneous observations (Wallyn et al., 2001). 

(3) A hard spectral component from $\sim$500 keV to several MeV with photon index of $\sim$0 were reported by several investigators (Hiller et al. 1970; Baker et al. 1973; Gruber $\&$ Ling, 1977). Specifically, HEAO-3 (Ling $\&$ Dermer 1991) observed a knee structure in 0.4 - 1 MeV region followed by a power law extending to $\>$3 MeV over a 50-day period in the spring of 1980. The knee structure was not observed by HEAO-3 six months earlier in the fall of 1979. 

(4) A final COMPTEL spectrum of the Crab pulsar/nebula emission in the 0.75-30 MeV range, using all available data of COMPTEL observations of the source and improved instrument response was recently presented by Kuiper et al. (2001). The spectrum indicates a possible broad spectral feature in the 0.7-3 MeV region followed by a power law with spectral index of $\sim$2 up to 30 MeV. The 0.7-3 MeV feature was also hinted in an earlier COMPTEL spectrum obtained by van der Meulen et al. (1998) but at lower fluxes.

The non-pulsed Crab nebula emission is generally interpreted as synchrotron radiation produced by relativistic electrons that were ejected by the pulsar, spiraling in the magnetic field of the nebula with strength of $\sim$10$^{-4}$G. 
Recent optical and X-ray images observed by HST (Hester et al. 2002), and Chandra (Weisskopf et al. 2000; Mori et al. 2003), respectively, revealed a highly dynamical structure of the system that consists of an inner ring, a torus, moving wisps, knots and a jet. X-ray spectra measured by XMM-Newton (Willingale et al. 2001) showed variation of the spectral shape across the nebula with the hardest spectral index seen close to the pulsar and softest in the outer nebula and the jet. 
Information about the gamma-ray (30 keV to 2 MeV) spectrum and variability need to be considered in the light of this revelation. The large body of information described above from the last three decades, provides some basis for this effort. However, the heterogeneous mixture of viewing periods, energies, and experiment characteristic make a confusing picture in the presence of source variability. Clearly a homogeneous and continuous set of observations is highly desirable to provide a clearer picture of the Crab's gamma-ray spectrum and variability. Questions that need to be addressed include: (1) Is the gamma-ray emission indeed variable as suggested by previous observations? Specifically, is the spectrum below 300 keV always a broken power law or does it occasionally change to a single power law? The different spectral components may be associated with the different features observed by HST, Chandra and XMM-Newton likely arising from different environments. (2) What are the ranges of the varying spectral parameters? What is the typical time scale for these changes? Is the time scale consistent with that seen in X-rays? (3) What is the gamma-ray spectrum between 300 keV and 2 MeV? Is it also variable? 

This paper addresses these issues using the JPL BATSE Earth Occultation database, EBOP (Ling et al. 1996, 2000), which provides a reasonably continuous monitor of the Crab's gamma-ray emission over the nine-and-half year mission. In this paper, we report results covering the first $\sim$1200-day period between 1991 and 1994 which were fully processed by EBOP in 1996 (Ling et al. 2000).  We have recently resumed processing the remaining six years of BATSE data covering the period from 1994 to the end of the CGRO mission in 2000. These results will be reported separately. A brief description of the EBOP database and technique is given in Section 2. Results produced by EBOP are presented in Section 3, and a discussion of the results is given in Section 4.

\section{EBOP Database}

The eight Large Area Detectors (LADs) of the BATSE experiment, each 50.8 cm diameter and 1.25 cm thick ($\sim2025$ cm$^2$ geometrical area), provided an unprecedented capability for independently monitoring the long-term behavior of gamma-ray sources with high spectral sensitivity in the energy range 0.02 - 1.8 MeV using the earth occultation technique.
Several previous publications have provided details of the experiment (Fishman et al. 1989), the BATSE earth occultation database and technique developed by the PI team at MSFC (Harmon et al. 2002), and the Enhanced BATSE Occultation Package (EBOP) developed by the JPL team (Ling et al. 1996, Ling et al. 2000).
Results presented in this paper were derived exclusively from EBOP. Its processed data products typically consist of daily fluxes for 64 known sources in 14 energy channels between 35 keV and 1.7 MeV, measured independently by each of the eight BATSE LADs. 
Each day, any given source, such as the Crab, was typically observed by two to four of the eight BATSE LADs with good sensitivity. These "source-viewing" LADs were selected using the criteria discussed by Ling et al. (1996, 2000).
Because the eight BATSE LADs were individually placed at different locations and oriented differently on the spacecraft, the background rates and the responses of the detectors to the source were distinctly different from one LAD to another, as were the associated source count rates.

A strength of the BATSE experiment and the EBOP system is that the source fluxes measured simultaneously by all source-viewing LADs on a given day, which were derived by unfolding the source count-rates through their corresponding detector response functions, can be directly compared, and must be consistent with one another. Achieving consistency requires that both the critical components in determining the source fluxes, namely, the source count-rates derived by the EBOP model, and the detector response matrices determined by the MSFC PI team, be correspondingly accurate. 

Fluxes for 604 days of observations of the Crab nebula between 1991 May 17 and 1994 October 17 included in this paper have passed the consistency test, as previously described by Ling et al. (1996, 2000). We exclude days with probability P($\chi^2, \nu$) $<$ 5\% from further consideration.

Figure 1 shows a sample of nine single-day spectra produced by EBOP. Panels 1-3 contain two simultaneous source-viewing LAD-spectra, panels 4-6 panels have three, and panels 7-9 have four, respectively.  The solid line is the best-fit single power-law model to the $n$ data points, where $n$ = number of LADs x 14 energy channels, using the standard analysis fitting program XSPEC (Arnaud 1996).  The best-fit model and parameters, as well as the reduced $\chi^2$ ($\chi^2$/$\nu$, where $\nu$ is the number of degrees of freedom) of the fit are 
also displayed in each panel. The level of consistency achieved in these samples, as reflected also by the goodness of the model fits, is a good representation of the 604 single-day spectra that have passed the EBOP consistency test. 

\section{Results}
\subsection{Flux Histories}

Figure 2 shows the flux histories, with 1-day resolution, in the 
35-100 keV, 100-200 keV, 200-300 keV, 300-400 keV, 400-700 keV, and 
700-1000 keV energy bands, respectively, covering the period between 1991 
May 24 (TJD 8400) and 1994 October 2 (TJD 9627).
The solid line in each panel represents the weighted mean of the 604 daily fluxes. Its value and standard error are also displayed in the upper right-hand corner in each panel. 
Key results shown in this figure are the following: 

(1) Positive fluxes were measured by BATSE in all six energy bands, with significances averaging ~$\sim31\sigma$, 20$\sigma$, 9.2$\sigma$, 4.5$\sigma$, 2.6$\sigma$, and 1.3$\sigma$, respectively, for single days.

(2) Fluxes shown in each of the three low-energy bands (1st, 2nd and 3rd panels) show significantly variations about their corresponding weighted mean value described above. Such variations are represented by the sum of the $\chi^2$ of individual daily fluxes about the weighted mean. The reduced $\chi^2$s for these three energy bands, with 603 degrees of freedom, are, 3.2, 2.0, and 1.3, respectively.   

(3) Data shown in the three high-energy bands, however, are more consistent with constant values. The reduced $\chi^2$ for these bands are 1.1, 1.1 1.0, respectively. 

The higher reduced $\chi^2$ in the three low-energy bands suggest intrinsic variability in the source emission in energies below 300 keV. In the next section, we present results of spectral variability in this energy region that were also observed by BATSE. 

\subsection{Spectra}
\subsubsection{Variable Broken Power-Law Spectra Below 300 keV}

Because of the statistical limitation of single-day spectra such as those shown in Figure 1, a simple power-law adequately characterizes most of the 604 spectra under study. However, as we build up statistics by summing data over many days (e.g. 4-13 days in a typical Viewing Period), a single power-law is no longer adequate in most cases, and a broken power-law is preferable. There were seventy-three Viewing Periods (VPs) that spanned the period between 1991 May 24 (TJD 8400) and 1994 October 2 (TJD 9627) (see Table 1) that have four or more single-day spectra that have passed the consistency test. In each VP, the spacecraft maintained a fixed orientation in inertial space. Consequently, the response to the Crab for each of the eight LADs, was the same throughout the VP. We produced, for each of the 64 sources in the EBOP catalog (see Ling et al. 2000), both single-day spectra such as those shown in Figure 1, as well as VP-average spectra. The latter were obtained by combining the source count-rates obtained for all the days in a VP which pass the consistency test, folded through each LAD response to the source for the VP, using XSPEC. 

Table 1 lists the pertinent information for the 73 VP spectra included in this report, including the VP number, the range of days and the number of ``clean" days included in the VP, the ``source-viewing" LADs and the best-fit parameters for both a single power-law or a broken power-law representation of the data. 

In Figure 3-a,b,c, we select 36 of the 73 VP spectra listed in Table 1 to illustrate key spectral features and variability. In each panel, the spectrum is plotted in the form of E$^2$ times Flux vs. Energy. Error bars plotted are of 1$\sigma$ significance, and upper limits are at the 2$\sigma$ level. The solid line is the best-fit broken power-law to the $n$ data points, where $n$ = 14 x number of ``source viewing" LADs as discussed in Section 2. The best-fit parameters are also shown in each panel. In the four panels of the middle column in each figure spectra measured simultaneously by all ``source-viewing" LADs during the VP are shown and compared to illustrate the level of consistency among the LADs that was achieved in these VP spectra, similar to that shown for the single-day spectra in Figure 1.
 
Key results shown in Table 1 and Figure 3 are summarized as follows:

(1) Sixty-one of the 73 VP spectra listed in Table 1 show lower reduced $\chi^2$ with a broken power-law model than the single power-law. 

(2) For twelve VPs, a single power-law ($\alpha =$ 2.13-2.20) fits essentially as well as a broken power-law. These are VP-9, 11, 42, 44, 218, 220, 231, 232.5, 305, 314, 319, and 322 (see panels 6, 23, 26, and 35 in Figure 3).

(3) The model parameters shown in Table 1 and Figure 3 indicate significant spectral variability from VP to VP. We compute the mean values and standard deviations of the three key broken power-law parameters, $\alpha_1$, $E_b$, and $\alpha_2$, averaged over the complete set of 73 VPs. They are 2.1 $\pm$ 0.1, 123 $\pm$ 42 keV, and 2.4 $\pm$ 0.2, respectively.

(4) The single or broken power-law can adequately fit only about half of the 73 VP spectra with acceptable $\chi^2$. The relatively poor fits shown in $\sim$33 of the 73 VP spectra (e.g.VP-5, 10, 12, 13, 16, 17, 18, 29, 39, 40, 42, 203, 203.3, 204, 207, 209, 218, 221, 226, 227, 228, 229.5, 232.5, 302.3, 303.2, 305, 310, 314, 322, 325, 332, 337, 338.5) are likely caused by the following effects.

(a) Short-term (single-day) spectral and flux variations shown in Figures 1 $\&$ 2, respectively due to intrinsic changes in the system preclude the possibility for a simple model (e.g. power-law or broken power-law) to adequately fit the long-term VP spectrum. As an illustration, Table 2 shows a comparison of the best-fit parameters for each of the eleven single-day spectra in VP-5 with those of the integrated VP spectrum. These results show that while a power-law model fits the single-day spectra adequately with spectral indices varying from 2.14$\pm$0.05 to 2.29$\pm$0.05, neither a single power-law (Table 2) nor a broken power-law (see Table 1) can fit the VP spectrum well. 

(b) A poor fit to the multiple-day VP spectrum compared to single-day spectrum, could be also due to the smaller errors associated with fluxes which make it more difficult for a simple model to fit the former spectrum than the latter. This is reflected to some extent by the results shown in Table 1 that twenty-two of the 33 VP spectra listed were averaged over eight or more days with better statistics compared to 11 spectra with less than eight days with relatively larger error bars.  

(c) The possible presence of a component above 300 keV that is suggested by many of the VP spectra (e.g. see Figure 3 panels 4, 5, 6, 8, 10, 11, 18, 21, 23). 

\subsubsection{Spectrum Above 300 keV}

Hints of excess high-energy flux (e.g. $>$300 keV) above the broken power-law model, as shown in many of the VP spectra, warrant a more careful study of the 300-1700 keV energy region. Because the high-energy fluxes above 300 keV (Figure 2, panels 4-6) appear to be quite stable over the 1200-day period, we group the 604 single-day spectra into three $\sim$400-day periods, namely TJDs 8400-8800, 8800-9200, and 9200-9628, respectively. Figure 4 shows a comparison of the three weighted average spectra of single-days for the three periods, plotted in photon flux units in the left panel (see also Ling et al. 2000 Figure 8 panel 11), and E$^2$ x Flux units in the right panel. The average fluxes for each of the three periods and their sum are shown in Table 3. There were 187 ``clean" single-day spectra in the first period (TJD 8400-8800), 197 in the second period (TJD 8800-9200), and 220 in the third period (TJD 9200-9628). The three spectra, which are nearly indistinguisible, can be best described as follows:

(1) From 35 to $\sim$670 keV, the spectrum has the shape of a broken power-law with spectral indices $\alpha_1$ = 2.10, $\alpha_2$ = 2.35 around the spectral break at $E_{b}$ = 112 keV. The extrapolation of the 1-10 keV power-law spectrum of (9.59$\pm$0.05)E$^{-2.108\pm0.006}$ photons cm$^{-2}$-s$^{-1}$-keV$^{-1}$ measured by XMM-Newton (Willingale et al., 2001-dashed line) is also shown in Figure 4 for comparison. The spectral index of 2.108 measured by XMM-Newton is consistent with the averaged index over the 73 VP spectra of 2.1 $\pm$ 0.1 below $\sim$100 keV measured by BATSE (see Section 3.2.1). The extrapolation of the XMM-Newton flux to $\sim$40 keV of 4 x 10$^{-3}$ photons cm$^{-2}$-s$^{-1}$-keV$^{-1}$ is also roughly consistent with the flux of $\sim$4.5 x 10$^{-3}$ photons cm$^{-2}$-s$^{-1}$-keV$^{-1}$, measured by BATSE (Figure 4 left panel).

(2) Above 670 keV, the spectrum becomes harder. The average of the three spectra in the 766-1104 keV bin is (4.67 $\pm$ 0.19) x 10$^{-6}$ photons$/$cm$^{-2}$-s$^{-1}$-keV$^{-1}$ ($\sim$25$\sigma$), and (2.26 $\pm$ 0.14) x 10$^{-6}$ photons$/$cm$^{-2}$-s$^{-1}$-keV$^{-1}$ ($\sim$16$\sigma$) in the 1104-1700 keV bin. They can be approximated by a power law with index of 1.75.

(3) There also appears to be a small excess of the 429-595 keV flux bin above the best-fit broken power-law model, in all three spectra. This energy band includes the various possible red-shifted annihilation features reported by several investigators in the past (Leventhal, MacCallum $\&$ Watts, 1977; Agrinier et al. 1990; Massaro et al. 1991). The estimated excesses, by subtracting the underlying power-law continuum from the fluxes measured in the 429-595 keV bin, are (1.8$\pm$1.4) x 10$^{-4}$, (3.9$\pm$1.4) x 10$^{-4}$, and (2.4$\pm$1.5) x 10$^{-4}$ photons cm$^{-2}$-s$^{-1}$-keV$^{-1}$, respectively, with a weighted average of (2.7$\pm$0.8) x 10$^{-4}$ photons cm$^{-2}$-s$^{-1}$-keV$^{-1}$.

Observations of the low-energy broken power-law of the Crab spectrum ($<$ 300 keV) had been reported by several investigators in the past (eg. Strickman et al. 1979; Jung 1989). The hardening of the spectrum $>$670 keV is new. If confirmed, it imposes new constraints on our understanding of the system. The apparent excess flux could conceivably be the result of systematic effects associated with the instrument's response. For example, an under-estimation of the response matrix above 670 keV would cause an over estimation of the high-energy source flux. But if the response matrix were indeed seriously erroneous, it would affect other source spectra as well. Figure 5 shows a direct comparison of the Crab spectrum with that of Cygnus X-1 for the same period, TJD 8800-9200. The two spectra are distinctly different. The Cygnus X-1 spectrum shows the standard Comptonized component below 300 keV plus a soft power-law tail above 300 keV, as reported by several investigators in the past (Ling et al. 1997; Philip et al. 1997; Ling 2001; McConnell et al. 2000, 2002). At $\sim$1 MeV, the Cygnus X-1 flux is about a factor of four lower than that of the Crab nebula, and is statistically consistent with the extrapolation of the power-law from low energy. This comparison suggests that the hardening of the three Crab spectra shown in Figure 4 (right panel) may indeed be real, and not caused by problems associated with instrument calibration.

Figure 6 compares the Crab's spectra from EBOP with those measured simultaneously by other CGRO experiments over differing time intervals: (a) VPs 1, 213 and 221 (top panel, BATSE, OSSE, COMPTEL and EGRET) reported by Much et al.(1996), which is a subset of, (b) 28 VPs shown in the COMPTEL list (Kuiper et al. 2001; except for VP-0, 36 and 724.5) for an averaged EBOP spectrum of 289 days (bottom panel). Included in this figure are also the COMPTEL spectrum reported by Kuiper et al. (2001), and EGRET spectrum by Fierro et al. (1998). In the top panel, the BATSE spectrum as analyzed by the PI team at MSFC, using a different earth-occultation analysis package with the same response matrix (Harmon et al. 2002) was also included in the paper by Much et al. (1996) for comparison. The two earth-occultation spectra are generally consistent with each other, but higher than the OSSE spectrum, especially above $\sim$300 keV. In particular, both earth-occultation spectra show consistent high fluxes in the $\sim$429-595 keV bin, and spectral hardening above $\sim$750 keV. The reason for the discrepancy with OSSE is unknown. 
In the lower panel, both EBOP and COMPTEL (Kuiper et al. 2001) spectra show similar slopes in the overlapping region between 750 keV and 1.5 MeV. The COMPTEL spectrum (Kuiper et al. 2001) shows a complex structures in the 0.7 - 3 MeV region. Its flux at $\sim$ 0.75-1 MeV is only lower than BATSE's by $\sim$6$\%$, suggesting that the "hard" spectral component around 1 MeV observed by BATSE could in fact be part of this complex structure observed by COMPTEL. The COMPTEL spectrum in the 3-25 MeV range, and EGRET (Nolan et al. 1993; Fierro et al. 1998) flux from 30-50 MeV are consistent with the extrapolation of the low-energy power-law spectrum (solid line). From $\sim$50 MeV to $\sim$100 MeV, the EGRET spectrum (Fierro et al. 1998) significantly softens, with an index $>$4. It finally flattens again to an index of $\sim$2 from $\sim$100 MeV to $\sim$1 GeV. 

\section{Discussion}

Highlights of the BATSE-EBOP results presented in this paper are summarized as follows:

$\bullet$ We observed significant flux (Figure 2) and spectral (Figure 3) variations in the total gamma-ray emission from the Crab nebula in the 35 - 300 keV energy region with time scale of days to weeks. The spectrum in this energy range, averaged over many days of typical viewing periods of the CGRO mission can be generally characterized by a variable broken power law (Figure 3 and Table 1). The average of the three broken power-law parameters, $\alpha_1$, $E_b$, and $\alpha_2$ over the complete set of 73 VPs under study are 2.1 $\pm$ 0.1, 123 $\pm$ 42 keV, and 2.4 $\pm$ 0.2, respectively. These averages include
contributions from twelve VP spectra (of 73 total) that can be fitted equally well with a single power-law.

$\bullet$ The high-energy fluxes above 300 keV, however, appear to be fairly constant (Figure 2). The long-term spectra, averaged over three 400-day periods, TJD 8400-8800, 8800-9200, and 9200-9628, respectively, are nearly indistinguishable (Figure 4). They can be well represented by a power law up to $\sim$670 keV that is consistent with the extrapolation of the broken power law from low-energy fluxes, and a harder spectral component from $\sim$670 keV to $\sim$1.7 MeV with a spectral index of $\sim$1.75 (Figure 4-right panel). The latter shape was also observed in the Crab spectrum measured by COMPTEL (Kuiper et al. 2001) as part of a complex structure in the 0.7-3 MeV range. In a direct comparison with the COMPTEL spectrum over approximately the same observational periods (Figure 6 lower panel), the flux reported by COMPTEL at $\sim$0.75-1 MeV is lower than that of BATSE by $\sim$6$\%$. Above 3 MeV, the extrapolation of the power-law continuum determined by the low-energy BATSE spectrum (solid line in Figure 6-lower panel) is consistent with fluxes measured by COMPTEL (Kuiper et al. 2001) in the 3-25 MeV range, and by EGRET's (Fierro et al. 1998) flux in the 30-50 MeV range. 

$\bullet$ BATSE also detected hints of excess flux in the 429-595 keV bin over the best-fit broken power-law model in the three 400$^{d}$- period spectra. This energy band includes possible contributions from features (variously attributed to red or blue-shifted positron annihilation) that have been reported in the past. The net excess flux averaged over the three spectra is estimated to be (2.7$\pm$0.8) x 10$^{-4}$ photons cm$^{-2}$-s$^{-1}$-keV$^{-1}$.

Gamma-ray emission between 30 keV and 2 MeV from the Crab nebula is generally thought to be produced by the interactions of electrons injected by the Crab pulsar into the surrounding synchrotron nebula. Details of the physical processes for producing the spectra and variability observed by BATSE, however, need to be interpreted in terms of what we have learned about the system today. Recent observations by HST (Hester et al.1995; Hester et al. 2002), ROSAT (van den Bergh and Pritchet, 1989), XMM-Newton (Willingale et al. 2001), and Chandra (Weisskopf et al. 2000; Mori et al. 2003) have provided us fascinating visible and X-ray images showing a highly complex and dynamic system that consists of an inner ring with bright knots, a thick torus with circular loops at two extremities, fast and slow moving wisps, and fast and slow moving jet and counter-jet, all embedded in the large synchrotron nebula. Several noteworthy features based on these images are:

1. The system is roughly cylindrically symmetrical about an axis from the southeast to the northwest, with the northwest axis pointing $\sim$30$^o$ toward us with respect to the plane of the sky (Hester et al. 1995; Weisskopf et al. 2000; Mori et al. 2003)

2. An inner X-ray ring observed by Chandra (Weisskopf et al. 2000) is thought to be caused by shocks associated with the pulsar wind hitting the slowly moving synchrotron-emitting plasma (Kennel \& Coroniti 1984). The ring has a semimajor axis of $\sim$0.14 pc, and a semiminor axis of $\sim$0.07 pc projected onto the sky. 

3. Approximately two dozen bright X-ray knots have been seen in the ring. These knots typically form and move within the ring while their intensity brightens and dimes, with occasional outbursts. Similar but much less conspicuous knots have also been seen at visible wavelengths. 

4. Moving wisps have been seen in both x rays by Chandra (Weisskopf et al.2000) and visible light by HST (Hester et al. 1995; 2002). The time interval between Chandra and HST observations (11 and 22 days, respectively) are comparable to the time scale of the variability shown by BATSE. The Chandra movie from their website showed that these wisps appear to emerge near the inner ring and move outward with velocity v$\sim$0.5c. These moving wisps form the basic fibrous texture of a thick outer torus whose semimajor axis is $\sim$0.37 pc, and semiminor axis is $\sim$0.18 pc. The torus has a minor radius of $\sim$0.08 pc. Wisps have also been seen (Hester et al. 2002) to emerge and move out into the inner nebula at a much lower speed of $\sim$0.03c resulting in a significantly less variable structures in the outer nebula compared to the dynamic effects seen between the inner ring and the torus. Large circular loop structures were also seen at the two extremities, similar to loops seen near the solar limb. HST data (Hester et al. 2002) suggested that these structures appear to be caused by magnetic field lines in the surrounding nebula that draped around and confined the expanding torus.

5. A bright jet-like structure in the southeast, and a dimmer counterjet in the northwest of the pulsar, approximately along the axis of symmetry, were also seen in both X-ray and visible wavelengths. The southeast jet appears narrower in the inner region near the pulsar, and more diffuse in the outer region (Weisskopf et al. 2000), and is highly dynamic, moving at a speed of $\sim$0.4c. It then slows down to $\sim$0.03c  as it pushes against the surrounding synchrotron nebula (Hester et al. 2002).

The Crab gamma-ray spectra presented in this paper correspond to the total time averaged emission spatially integrated over entire nebula, including both the pulsar and the surrounding regions discussed above. The phase-averaged spectrum of the pulsar measured by HEAO-3 (Mahoney, Ling $\&$ Jacobson, 1984), which can be described by a single power law 1.04 x 10$^{-4}$(E/100)$^{-2.14}$ photons cm$^{-2}$-s$^{-1}$-keV$^{-1}$, accounts $\sim$15\% and $\sim$18\% to the observed flux by BATSE at 100 keV and 1 MeV, respectively. At 10 keV, the pulsed flux (Kuiper et al. 2001) accounts to $\sim$10\% of the total emission. The rest must be due to contributions from other parts of the system such as those described above. The broad-band pulsed and nebula spectra, from soft x-rays to TeV gamma-rays, compiled by Kuiper et al. (2001) show that the pulsed spectrum in the 1 to $\sim$10 keV range a relatively hard, with spectral index of $\sim$1.6, compared to $\sim$2 for 10 - $\sim$500 keV component. The nebula spectrum, on the other hand, can be characterized by a single power-law with spectral index of $\sim$2.15 for the full energy range from 1 keV to $\sim$1 MeV. 
Willingale et al. (2001) showed that the total Crab spectrum (nebula \& pulsed)in the 1-10 keV region measured by XMM-Newton can be described as (9.59$\pm$0.05)E$^{-2.108\pm0.006}$ photons cm$^{-2}$-s$^{-1}$-keV$^{-1}$. The close resemblance of the spectral index of 2.108 for the total emission compared to that the nebula component of 2.15 discussed above, suggests that the nebula emission is the dominant component of the total emission in the 1-10 keV range as expected. Furthermore, the spectral index of 2.108 in the 1-10 keV range measured by XMM-Newtron is also consistent with the averaged index over the 73 VP spectra of 2.1 $\pm$ 0.1 from 35 to $\sim$100 keV measured by BATSE (see Section 3.2.1). The extrapolation of the XMM-Newton flux to $\sim$40 keV of 4 x 10$^{-3}$ photons cm$^{-2}$-s$^{-1}$-keV$^{-1}$ is consistent with the flux of $\sim$4.5 x 10$^{-3}$ photons cm$^{-2}$-s$^{-1}$-keV$^{-1}$, shown by the three 400-day spectra measured by BATSE (see Figure 4 left panel). While the spectral index for the total X-ray nebula emission measured by XMM-Newton is 2.1, Willingale et al. (2001) showed significant spectral variations across the nebula. The hardest spectrum was observed near the pulsar with an index of 1.6 $\pm$ 0.03. This is consistent with the pulsed spectrum shown in the paper by Kuiper et al (2001).  For the torus, it was 1.8 $\pm$ 0.006 compared to 1.9 measured by Chandra (Mori et al. 2003). The spectra for the jet and outer nebula are softer, with indices 2.10 $\pm$ 0.013 and 2.34 $\pm$ 0.005, respectively. The corresponding indices observed by Chandra (Mori et al. 2003) are 2.0-2.1 and $\sim$3.0, respectively. The fact that both the X-ray (1-10 keV) and hard X-ray (35-100 keV) spectra for the total emission (nebula \& pulsed) measured by XMM-Newton and BATSE, respectively, have the same spectral indices and flux normalization at $\sim$40 keV suggests that contributions from various regions of the system seen in 1-10 keV may be extended with the same ratio to $\sim$100 keV. The softest component seen by BATSE is in the $\sim$100-700 keV range with an index of 2.34. Is it dominated by emission from the outer nebula as shown in the x-ray data? Similarly, the hardest spectral component observed by BATSE is in the 0.7-1.7 MeV range with an index of 1.75. Is this component associated possibly with a broad spectral feature in the 0.7-3 MeV region observed by COMPTEL? If so, what is the mechanism for producing such a feature? 

The BATSE data, together with the recent x-ray observations by XMM-Newton (Willingale et al 2001) and Chandra (Weisskopt et al. 2000; Mori et al. 2003), and gamma-ray observations by COMPTEL (Kuiper et al 2001), EGRET (Fierro et al. 1998) and OSSE (Ulmer et al. 1995), should stimulate future theoretical modeling of the magnetic structure, particle transport, and gamma-ray emission processes in the Crab to address several fundamental questions about the system.

(1) What are the time and spatially dependent electron spectra at various regions of the system, such as the pulsar, torus, rings, jets and outer nebula. Do positrons play any significant role in the electron dynamics, and do their annihilation photons contribute significantly to the emission?

(2) What are the resulting synchrotron gamma-ray spectra produced in these regions, and how do they compare with the spatially integrated and time-dependent gamma-ray spectra observed? 

(3) What, if any, role do heavy ions play in the pulsar and nebula plasma?  Is the broad spectral structure indicated by COMPTEL and BATSE around 1 MeV related to ion, pair production or annihilation effects? 

Finally, in addition to the theoretical work, we hope future high spatial and high-spectral resolution ($\sim$few arcseconds spatial, 2-3 keV spectral) gamma-ray observations with new experiments such as RHESSI (Lin et al. 2003) and INTEGRAL will illuminate some of these questions. 

\acknowledgements
	We wish to thank Gerald Fishman, Alan Harmon and Mark Finger of the BATSE team for their support of the BATSE Earth Occultation investigation effort at JPL over the years. We appreciate the valuable comments made by an anonymous referee,  Wm. Mahoney and K. Mori on the manuscript. We are also grateful for the significant contributions made by undergraduate students Juan Estrella and Zachary Medin for processing the data. The work described in this paper was carried out by the Jet Propulsion Laboratory, under the contract with the National Aeronautics and Space Administration.

\clearpage
FIGURE CAPTIONS

\figcaption{A sample of nine single-day spectra were selected to illustrate the level of consistency among all source-viewing LADs that has been achieved by EBOP on the daily basis. Panels 1-3 contain two simultaneous source-viewing LAD-spectra, panels 4-6 panels have three, and panele 7-9 have four, respectively.  The solid line is the best-fit single power-law model 
to the $n$ data points, where $n$ = number of LADs x 14 energy channels, using the standard analysis fitting program XSPEC (Arnaud 1996).  The best-fit model and parameters, as well as reduced $\chi^2$ ($\chi^2$/$\nu$, where $\nu$
is the number of degrees of freedom) of the fit are 
also displayed in each panel. The level of consistency of gamma-ray fluxes measured simultaneously by the source-viewing LADs are typically reflected by the goodness of the model fits,}

\figcaption{Flux histories, with 1-day resolution, in the 
35-100 keV, 100-200 keV, 200-300 keV, 300-400 keV, 400-700 keV, and 700-1000 keV energy bands, respectively, covering the period between 1991 
May 24 (TJD 8400) and 1994 October 2 (TJD 9627).
The solid line in each panel represents the weighted mean of the 604 daily fluxes. Its value and error are also displayed in the upper right-hand corner in each panel. Positive fluxes were measured by BATSE in all six energy bands with significance averaging ~$\sim31\sigma$, 20$\sigma$, 9.2$\sigma$, 4.5$\sigma$, 2.6$\sigma$, and 1.3$\sigma$, respectively. for single days. The reduced $\chi^2$ for fitting a constant value to the 604 daily fluxes in three low-energy bands  (panels 1, 2 and 3) are, 3.2, 2.0, and 1.3, respectively, suggesting intrinsic variability of source emission below 300 keV. Fluxes above 300 keV (panels 4-6), however, are more consistent with constant values. The reduced $\chi^2$ for the three high-energy bands are 1.1, 1.1 1.0, respectively.}

\figcaption{Thirty-six of the 73 VP spectra listed in Table 1, with 12 in each of the three figures 3-a, 3-b, and 3-c, were shown to illustrate the range of variability of the spectral parameters as observed by BATSE. In each panel, the spectrum is plotted in the form of E$^2$ x Flux vs. Energy. Error bars plotted are of one $\sigma$ significance, and upper limits are at the 2$\sigma$ level. The solid line is the best-fit broken power-law to the $n$ data points, where $n$ = 14 x number of "source viewing" LADs discussed in Section 2. The best-fit parameters are also shown in each panel. In the four panels of the middle column in each figure (e.g. panels 2, 5, 8, 11 in Figure 3a, panels 14, 17, 20, 23 in Figure 3b, and panels 26, 29, 32, 35 in Figure 3c), spectra measured simultaneously by all ``source-viewing" LADs during the VP are shown and compared to illustrate the level of consistency among the LADs that was achieved in these VP spectra, similar to that shown for the single-day spectra in Figure 1. While most of the VP spectra can be best-fitted with a broken power-law model, twelve of the 73 VP spectra as illustrated in panel 6 in Figure 3a, panel 23 in Figure 3b, and panels 26 and 35 in Figure 3c, can be fitted equally well with a single power-law.}

\figcaption {Three long-term weighted average spectra integrated over three separate 400${^d}$-periods, TJD 8400-8800, 8800-9200, and 9200-9628, respectively, are nearly indistinguishable (Figure 4-a,b). The extrapolation of the 1-10 keV power-law spectrum measured by XMM-Newton (Willingale et al., 2001-dashed line) is also included in Figure 4-a,b for comparison. The three BATSE spectra are consistent with a broken power law from 35 to $\sim$670 keV,with spectral indices $\alpha_1$ = 2.10, $\alpha_2$ = 2.35, about the spectral break energy at $E_{b}$ = 112 keV. Above 670 keV, the spectra become harder with a spectral index of 1.75. There appears also a small excess of the 429-595 keV flux bin above the best-fit broken power-law model, in all three spectra with an average flux of (2.7$\pm$0.8) x 10$^{-4}$ photons cm$^{-2}$-s$^{-1}$-keV$^{-1}$.}

\figcaption{A direct comparison of the Crab and Cygnus X-1 spectrum measured in the same period TJD 8800-9200 is shown. The two spectra are distinctly different. Cygnus X-1 spectrum shows the standard Comptonized component below 300 keV plus a soft power-law tail above 300 keV as reported by several investigators in the past (Ling et al. 1997; Philip et al. 1997; McConnell et al. 2000, 2002). At $~$1 MeV, the Cygnus X-1 flux is about a factor of four lower than that of the Crab nebula, and is statistically consistent with the extrapolation of the power-law from low energy.  This comparison suggests that the hardening of the three Crab spectra shown in Figure 5b may indeed be real, and it is not caused by problems associated with the instrument calibration.}

\figcaption{A direct comparison of the Crab's spectra from EBOP with those measured simultaneously by other CGRO experiments over differing time intervals: (a) VPs 1, 213 and 221 (top panel) reported by Much et al.(1996), which is a subset of (b) 28 VPs shown in the COMPTEL list (Kuiper et al. 2001; except for VP-0, 36 and 724.5) for an averaged EBOP spectrum of 289 single-day spectra(bottom panel). Included in the bottom figure are also the COMPTEL spectrum reported by Kuiper et al. (2001), and EGRET spectrum by Fierro et al. (1998). In the top panel, the BATSE/EBOP spectrum is generally consistent with that analyzed by the PI team at MSFC using a different analysis package (Harmon et al. 2002), and with the COMPTEL spectrum in overlapping regions. Both earth-occultation spectra are higher than the OSSE spectrum. The reason for the discrepancy is unknown. 
In the lower panel, both EBOP and COMPTEL (Kuiper et al. 2001) spectra show similar slopes in the overlapping region between 750 keV and 1.5 MeV. The COMPTEL spectrum (Kuiper et al. 2001) shows a complex structures in the 0.7 - 3 MeV region. Its flux at $\sim$0.75-1 MeV is only lower than BATSE's by $\sim$6$\%$, suggesting that the "hard" spectral component around 1 MeV observed by BATSE could in fact be part of this complex structure observed by COMPTEL. 
The COMPTEL spectrum in the 3-25 MeV range (Kuiper et al. 2001), and EGRET (Fierro et al. 1998) flux from 30-50 MeV are consistent with the extrapolation of the low-energy power-law spectrum (solid line) measured by BATSE. From $\sim$50 MeV to $\sim$100 MeV, the EGRET spectrum (Fierro et al. 1998) significantly softens, with an index $>$4. It finally flattens again to an index of $\sim$2 from $\sim$100 MeV to $\sim$1 GeV.} 

\newpage
\plotone{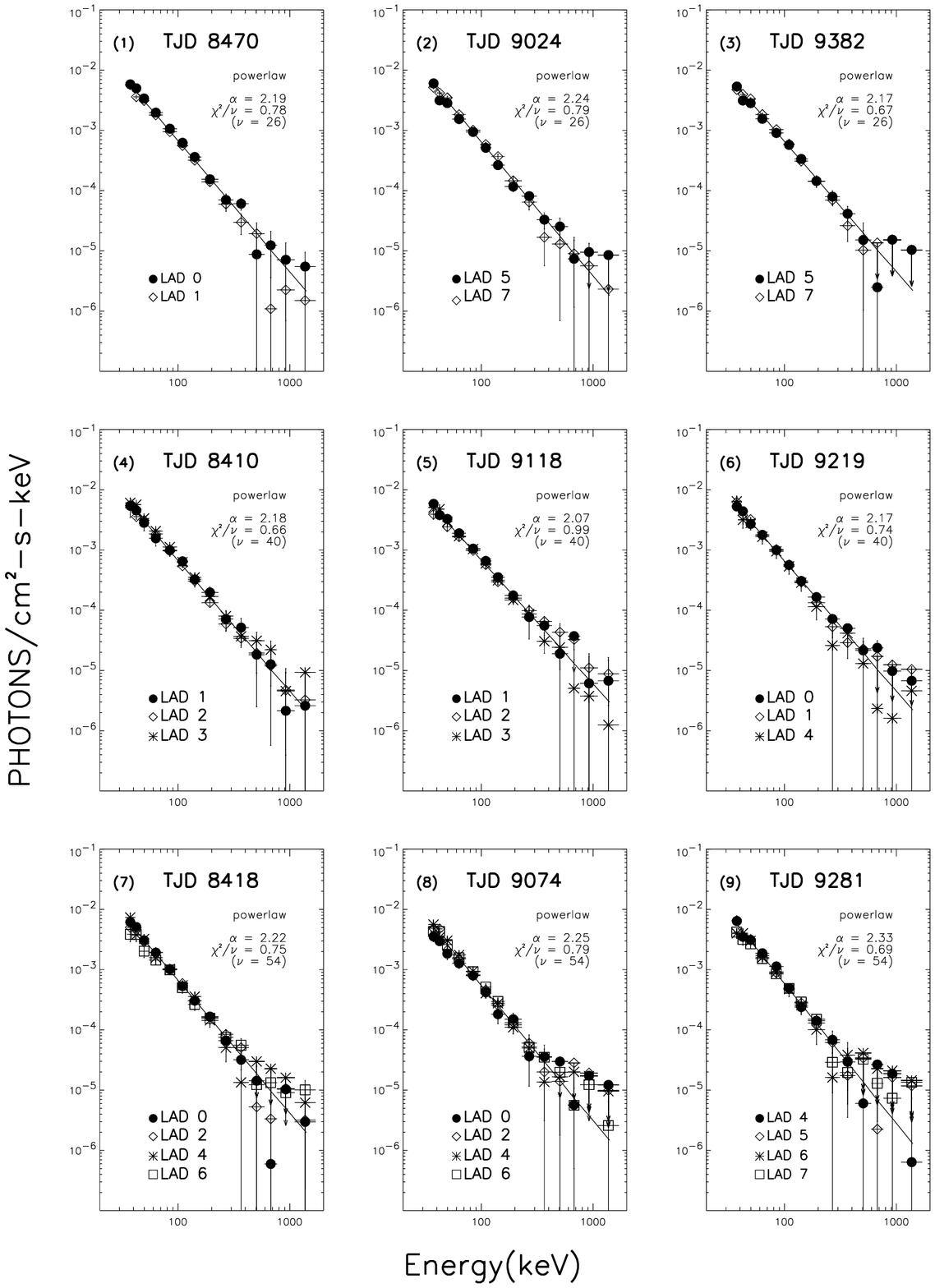}
\newpage
\epsscale{.90}
\plotone{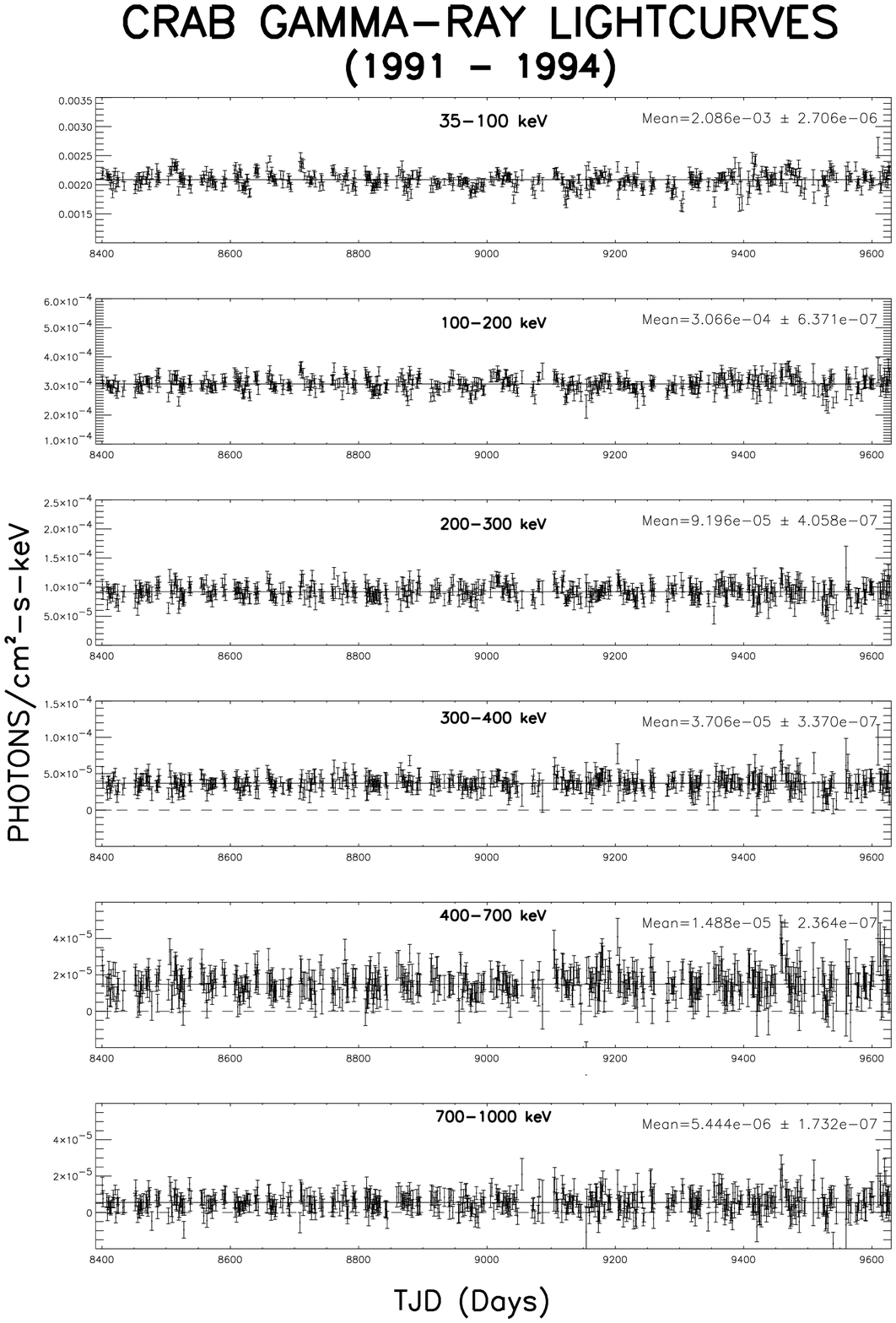}
\newpage
\epsscale{1.0}
\plotone{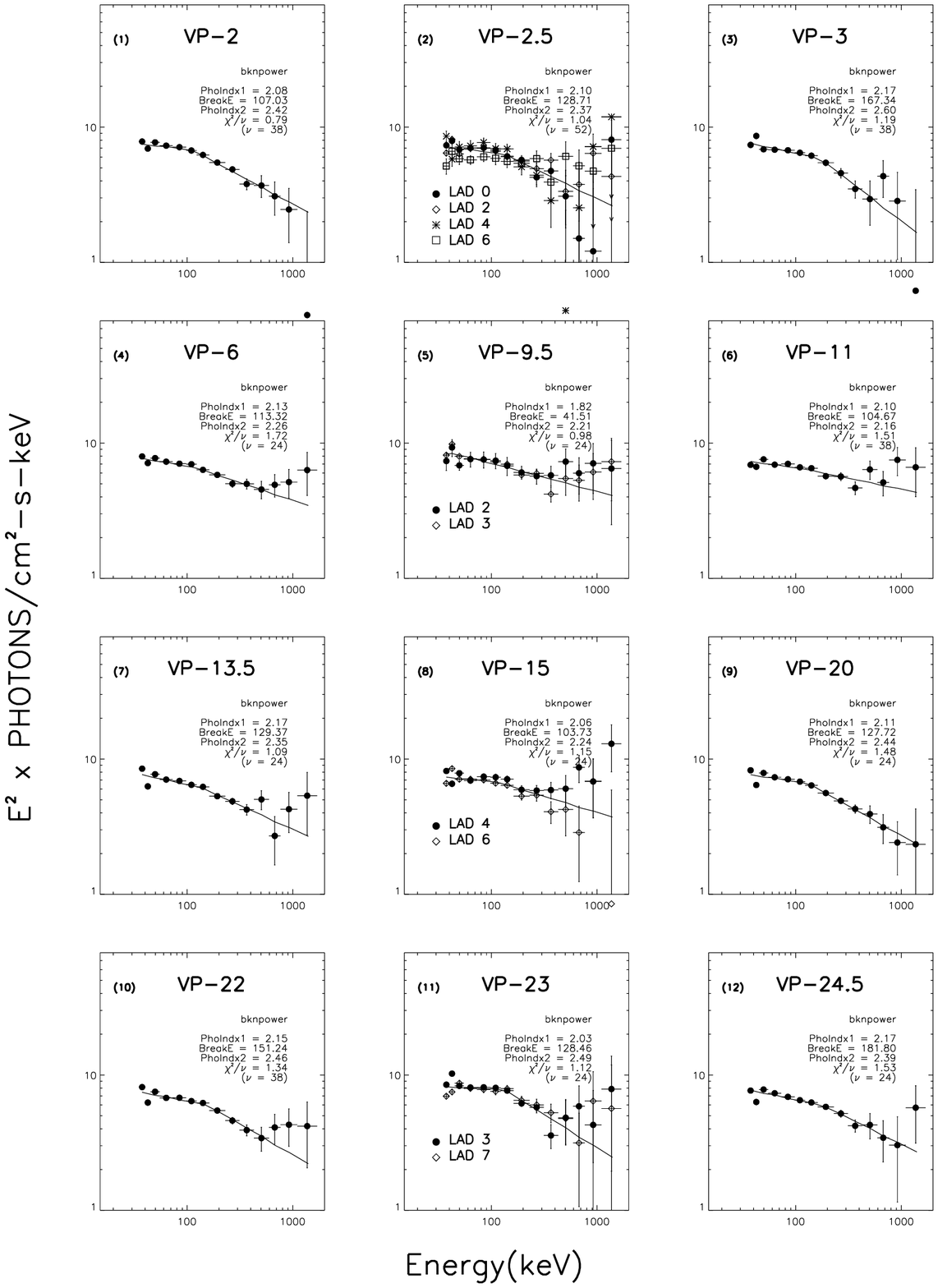}
\newpage
\plotone{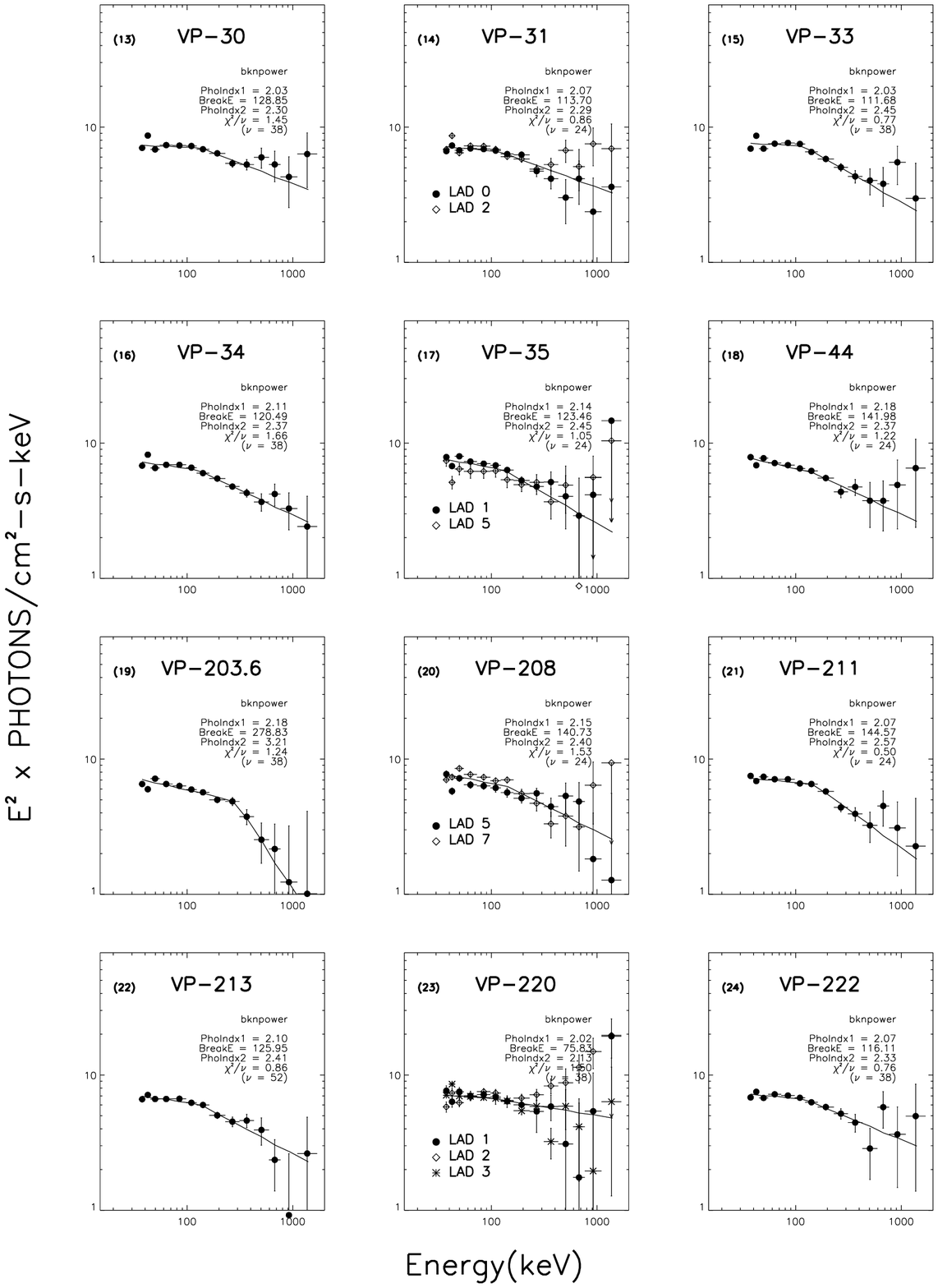}
\newpage
\plotone{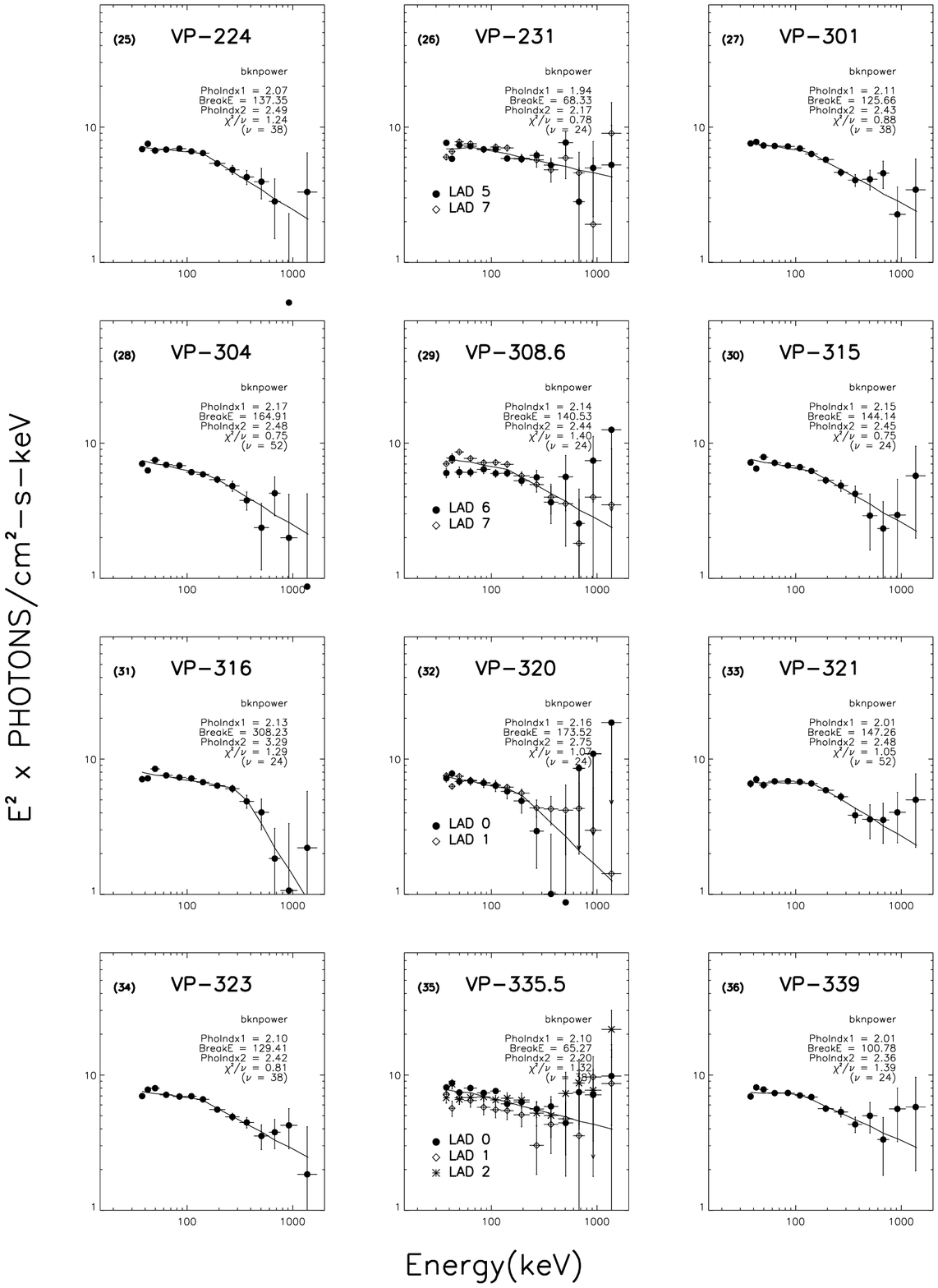}
\newpage
\plotone{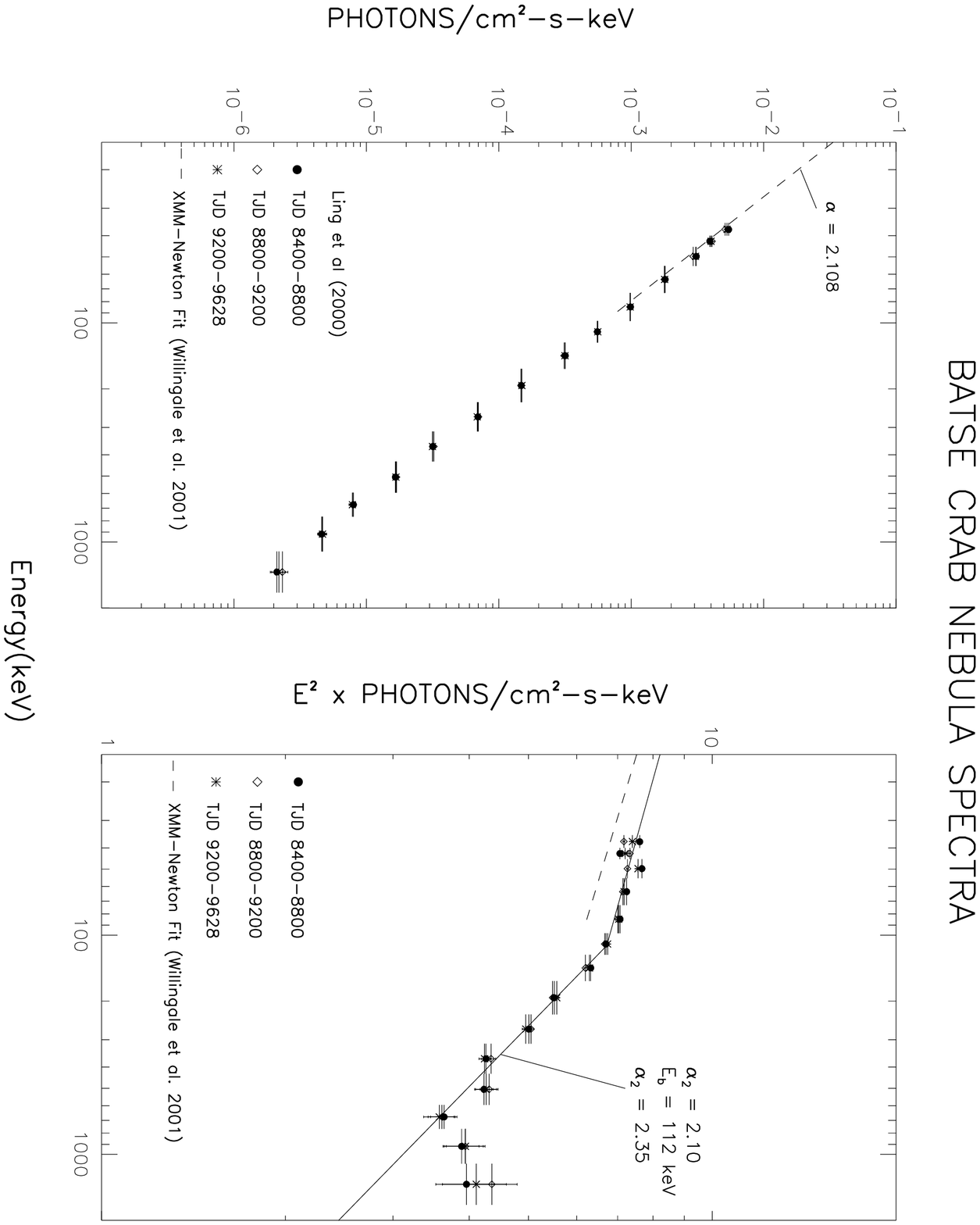}
\newpage
\plotone{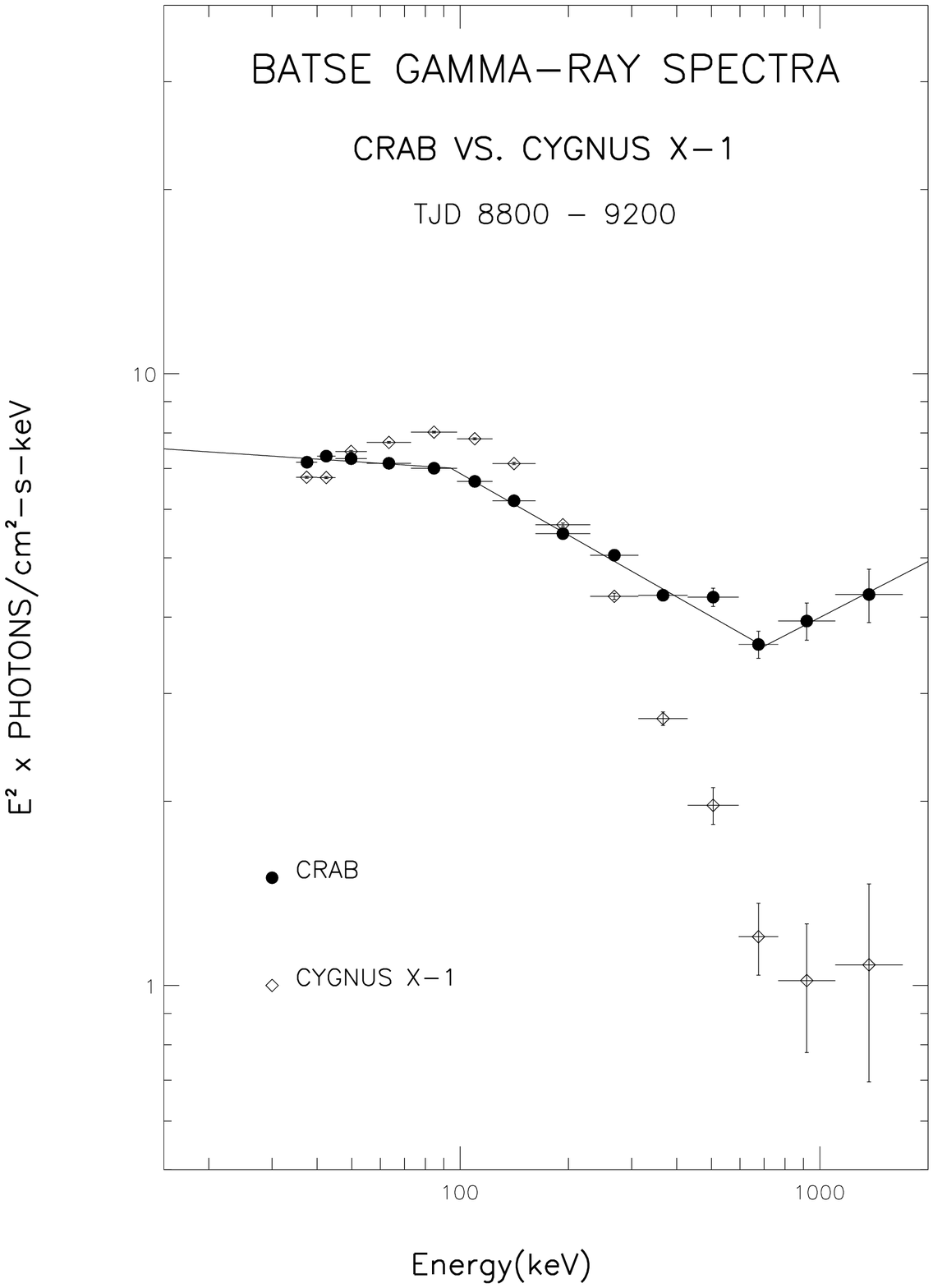}
\newpage
\plotone{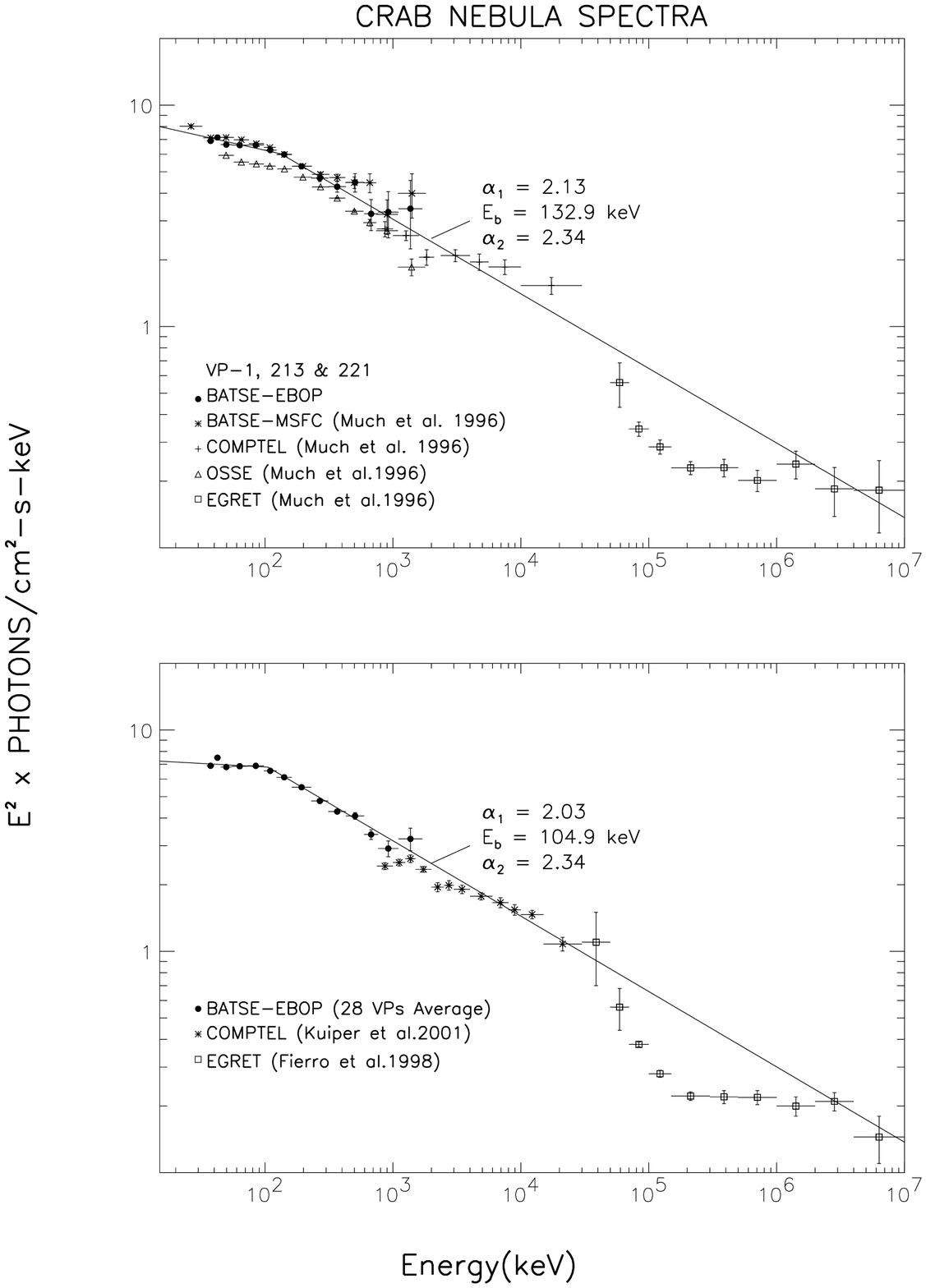}

\clearpage
\pagestyle{empty}
\begin{center}
\begin{deluxetable}{ccccccccccccccc}

\rotate

\tabletypesize{\tiny}

\tablewidth{660pt}


\tablenum{1}

\tablehead{
\multicolumn{5}{c}{TJD Included in VP \# of Clean} & 
\multicolumn{4}{c}{Single Power-law} &
\multicolumn{6}{c}{Broken Power laws} \\
\colhead{VP} & \colhead{TJD} & \colhead{Spectrum} & \colhead{Days} & 
\colhead{LADs} & \colhead{$\nu$} & \colhead{$\alpha$} & \colhead{$\chi^2$} & 
\colhead{$\chi^2/\nu$} & \colhead{$\nu$} & \colhead{$\alpha_1$} & \colhead{E$_b$} & 
\colhead{$\alpha_2$} & \colhead{$\chi^2$} & \colhead{$\chi^2/\nu$} \\
\colhead{} & \colhead{} & \colhead{} & \colhead{in VP} & \colhead{} & \colhead{} & 
\colhead{} & \colhead{} & \colhead{} & \colhead{} & \colhead{} & \colhead{} & 
\colhead{} & \colhead{} & \colhead{} 
} 

\startdata
2.0 & 8407-8414 & 8407-8414 & 8 & 1,2,3 & 40 & 2.20  $\pm$  0.01 & 78.8 & 1.97 & 38 & 2.08  $\pm$  0.03 & 107.0  $\pm$  13.7 & 2.42  $\pm$  0.05 & 30.2 & 0.79 \\
2.5 & 8416-8421 &  8416-8418,8421 & 4 & 0,2,4,6 & 54 & 2.19  $\pm$   0.02  & 82.4 & 1.53 & 52 & 2.10  $\pm$  0.04 & 128.7  $\pm$  37.1 & 2.37  $\pm$  0.09 & 54.0 & 1.04 \\
3.0 & 8423-8434 & 8423-8425,8432,8434 & 5 & 1,2,3 & 40 & 2.24   $\pm$  0.02   & 58.0 & 1.45 & 38 & 2.17  $\pm$  0.03 & 167.3  $\pm$  48.4 &  2.60  $\pm$  0.02 & 45.3 & 1.19 \\
5.0 & 8450-8462 & 8451-8453,8455-8462 & 11 & 1,3,5,7 & 54 & 2.21  $\pm$   0.01  & 172.6 & 3.20 & 52 & 2.12  $\pm$  0.03 & 123.3  $\pm$  28.5 & 2.42  $\pm$  0.09 & 145.5 & 2.80 \\
6.0 & 8464-8475 & 8464,8468-8473 & 8 & 0,1 & 26 & 2.18  $\pm$  0.01  & 46.6 & 1.79 & 24 & 2.13  $\pm$  0.03  & 113.3  $\pm$  43.5 & 2.26  $\pm$  0.06 & 41.2 & 1.72 \\
7.5 & 8484-8489 & 8484-8489 & 6 & 3,5,7 & 40 & 2.19 $\pm$  0.01  & 68.9 & 1.72 & 38 & 2.15 $\pm$ 0.03 & 139.1 $\pm$ 68.3 & 2.30 $\pm$ 0.09 & 64.3 & 1.69 \\
9.0 & 8505-8510 & 8505,8507-8509 & 4 & 4,5 & 26 & 2.22 $\pm$ 0.02 & 39.2 & 1.51 & 24 & 2.18 $\pm$ 0.06  & 89.4 $\pm$ 89.9 & 2.26 $\pm$ 0.07 & 38.3 & 1.60 \\
9.5 & 8512-8517 & 8512-8517 & 6 & 2,3 & 26 & 2.16 $\pm$ 0.01 & 42.8 & 1.65 & 24 & 1.82 $\pm$ 0.02  & 41.5 $\pm$ 8.3 & 2.21 $\pm$ 0.02 & 23.6 & 0.98 \\
10.0 & 8519-8531 &  8519-8528 & 10 & 4,5 & 26 & 2.24 $\pm$  0.01 & 96.5 & 3.71 & 24 & 2.15  $\pm$ 0.02  & 146.3 $\pm$ 24.5 & 2.60 $\pm$ 0.13 & 63.4 & 2.64 \\
11.0 & 8533-8545 & 8535-8538 & 4 & 5,6,7 & 40 & 2.13 $\pm$ 0.02 & 57.9 & 1.45 & 38 & 2.10 $\pm$ 0.05 & 105 $\pm$ 149  & 2.16 $\pm$ 0.08 & 57.4 & 1.51 \\
12.0 & 8547-8559 & 8554-8556, 8558-8559 & 5 & 1,5 & 26 &  2.13 $\pm$ 0.01  & 68.5 & 2.63 & 24 & 1.93 $\pm$ 0.09  & 61.5 $\pm$ 14.8 & 2.21 $\pm$ 0.03 & 53.0 & 2.21 \\
13.0 & 8561-8566 & 8562-8563, 8565-8566 & 4 & 3,5,7 & 40 &  2.22 $\pm$ 0.02   & 102.1 & 2.55 & 38 & 2.12 $\pm$ 0.03 & 145.8 $\pm$ 26.8 & 2.62 $\pm$ 0.15  & 73.6 & 1.94 \\
13.5 & 8568-8573 & 8568-8573 & 6 & 4,5 & 26 & 2.22 $\pm$ 0.01 & 33.4 & 1.28 & 24 & 2.17 $\pm$ 0.03   & 129.4 $\pm$ 47.9 &  2.35 $\pm$ 0.09  & 26.2 & 1.09 \\
15.0 & 8589-8601 & 8589-8592 & 4 & 4,6 & 26 & 2.13 $\pm$ 0.02  & 33.0 & 1.27 & 24 & 2.06 $\pm$ 0.04  & 103.7 $\pm$ 45.2 & 2.24 $\pm$ 0.07 & 27.6 & 1.15 \\
16.0 & 8603-8616 & 8606-8616 & 11 & 3,5,7 & 40 &  2.22 $\pm$ 0.01  & 128.0 & 3.20 & 38 & 2.13 $\pm$ 0.02 & 134.3 $\pm$ 21.9 & 2.48 $\pm$ 0.08 & 89.3 & 2.35 \\
17.0 & 8618-8630 & 8618-8627, 8629-8630 & 12 & 4,5,6,7 & 54 & 2.21 $\pm$ 0.01 & 137.3 & 2.54 & 52 & 2.14 $\pm$ 0.02 & 152.9 $\pm$ 36.1 & 2.38 $\pm$ 0.07 & 117.5 & 2.26 \\
18.0 & 8632-8643 & 8638-8643 & 6 & 4,5 & 26 & 2.21 $\pm$ 0.01 & 53.6 & 2.06 & 24 & 2.17 $\pm$ 0.03 & 123.6 $\pm$ 51.2 & 2.33 $\pm$ 0.07 & 49.5 & 2.06 \\
20.0 & 8659-8671 & 8659-8671 & 13 & 1,5 & 26 & 2.19 $\pm$ 0.01  & 90.3 & 3.47 & 24 & 2.11 $\pm$ 0.02 & 127.7 $\pm$ 16.6 & 2.44 $\pm$ 0.06 & 35.4 & 1.48 \\
22.0 & 8687-8696 &  8689-8694 & 6 & 0,4,5 & 40 & 2.22 $\pm$ 0.01 & 66.0 & 1.65 & 38 & 2.15 $\pm$ 0.02 & 151.3 $\pm$ 34.9 & 2.46 $\pm$ 0.10 & 51.0 & 1.34 \\
23.0 & 8701-8713 & 8708-8711, 8713 & 5 & 3,7 & 26 & 2.15 $\pm$ 0.02  & 59.3 & 2.28 & 24 & 2.03 $\pm$ 0.03 & 128.5 $\pm$ 21.5 & 2.49 $\pm$ 0.11 & 26.9 & 1.12 \\
24.5 & 8722-8727 & 8722-8727 & 6 & 5,7 & 26 & 2.19 $\pm$ 0.01 & 42.3 & 1.63 & 24 & 2.17 $\pm$ 0.02 & 181.8 $\pm$ 89.0 & 2.39 $\pm$ 0.20 & 36.8 & 1.53 \\
29.0 & 8757-8776 & 8758-8761, 8763, 8767, 8770, 8774, 8776 & 9 & 0,2 & 26 & 2.13 $\pm$ 0.01   & 123.6 & 4.75 & 24 & 1.97 $\pm$ 0.03  & 84.3 $\pm$ 10.0 & 2.33 $\pm$ 0.04 & 59.5 & 2.48 \\
30.0 & 8778-8783 & 8778-8783 & 6 & 0,2,3 & 40 & 2.11  $\pm$ 0.01   & 72.2 & 1.81 & 38 & 2.03 $\pm$ 0.03 & 128.8 $\pm$ 29.9 & 2.30 $\pm$ 0.08 & 55.1 & 1.45 \\
31.0 & 8785-8797 & 8790-8795 & 6 & 0,2 & 26 & 2.14 $\pm$ 0.01  & 31.5 & 1.21 & 24 & 2.07 $\pm$ 0.03  & 113.7 $\pm$ 33.2 & 2.29 $\pm$ 0.07 & 20.7 & 0.86 \\
33.0 & 8806-8818 & 8810-8818 & 9 & 0,2,3 & 40 & 2.16 $\pm$ 0.01 & 71.8 & 1.80 & 38 & 2.03 $\pm$ 0.03 & 111.7 $\pm$ 15.8 & 2.45 $\pm$ 0.08 & 29.1 & 0.77 \\
34.0 & 8820-8839 & 8820-8833 & 14 & 0,2,3 & 40 & 2.19 $\pm$ 0.01 & 92.8 & 2.32 & 38 & 2.11 $\pm$ 0.02 & 120.5 $\pm$ 21.0 & 2.37 $\pm$ 0.05 & 63.1 & 1.66 \\
35.0 & 8841-8844 & 8841-8844 & 4 & 1,5 & 26 & 2.22  $\pm$ 0.02   & 34.4 & 1.32 & 24 &  2.14 $\pm$ 0.04  & 123.5 $\pm$ 41.2  & 2.45 $\pm$ 0.14 & 25.2 & 1.05 \\
39.0 & 8867-8881 & 8867-8872, 8874-8881 & 14 & 0,2,6 & 40 & 2.14 $\pm$ 0.01   & 151.1 & 3.78 & 38 & 1.99 $\pm$ 0.03 &  88.9 $\pm$ 12.6 & 2.28 $\pm$ 0.03 & 104.5 & 2.75 \\
40.0 & 8883-8902 & 8883-8884, 8890-8895 & 8 & 4,6 & 26 & 2.15 $\pm$ 0.01 & 87.4 & 3.36 & 24 & 2.05 $\pm$ 0.03 & 119.3 $\pm$ 23.7 & 2.36 $\pm$ 0.08 & 61.6 & 2.57 \\
42.0 & 8911-8923 & 8912-8913, 8915, 8917, 8919, 8921-8923 & 8 & 1,5 & 26 & 2.16 $\pm$ 0.01 & 57.1 & 2.20 & 24 & 2.07 $\pm$ 0.07 & 53.3 $\pm$ 24.9 & 2.20 $\pm$ 0.03 & 52.4 & 2.18 \\
44.0 & 8930-8942 & 8933, 8940-8942 & 4 & 4,6 & 26 &  2.21 $\pm$ 0.02 & 32.7 & 1.26 & 24 & 2.18 $\pm$ 0.03 & 142.0 $\pm$ 79.2 & 2.37 $\pm$ 0.16 & 29.4 & 1.22 \\
203.0 & 8958-8963 & 8958-8959, 8961-8963 & 5 & 4,5,7 & 40 &  2.15 $\pm$  0.02  & 92.8 & 2.32 & 38 &  2.02 $\pm$ 0.05 & 86.0 $\pm$ 20.7 & 2.28 $\pm$ 0.05 & 77.4 & 2.04 \\
203.3 & 8965-8970 & 8965-8970 & 6 & 4,5,7 & 40 &  2.17 $\pm$  0.01 & 112.9 & 2.82 & 38 & 2.05 $\pm$ 0.03 & 90.1 $\pm$ 15.3 & 2.31 $\pm$ 0.04 & 84.3 & 2.22 \\
203.6 & 8972-8977 & 8973-8977 & 5 & 4,5,7 & 40 & 2.23 $\pm$ 0.02 & 63.9 & 1.60 & 38 & 2.18 $\pm$ 0.02 & 278.9 $\pm$ 74.7 & 3.21 $\pm$ 0.67 & 47.0 & 1.24 \\
204.0 & 8979-8984 & 8980-8984 & 5 & 5,6,7 & 40 & 2.21 $\pm$ 0.02 & 88.3 & 2.21 & 38 &  2.14 $\pm$ 0.03 & 163.8 $\pm$ 38.3 & 2.58 $\pm$ 0.17 & 70.2 & 1.85 \\
207.0 & 9000-9019 & 9005, 9007, 9010, 9014-9019 & 9 & 5,7 & 26 & 2.10 $\pm$ 0.01 & 63.4 & 2.44 & 24 &  2.03 $\pm$ 0.03 & 108.0 $\pm$ 26.1 & 2.22 $\pm$ 0.05 & 47.0 & 1.96 \\
208.0 & 9021-9026 & 9023-9026 & 4 & 5,7 & 26 & 2.21 $\pm$ 0.02 & 43.5 & 1.67 & 24 & 2.15 $\pm$ 0.03 & 140.7 $\pm$ 57.7 & 2.40 $\pm$ 0.13 & 36.8 & 1.53 \\
209.0 & 9028-9039 & 9028-9036, 9039 & 10 & 3,7 & 26 & 2.17 $\pm$ 0.01  & 124.9 & 4.80 & 24 &  2.10 $\pm$ 0.02 & 146.3 $\pm$ 24.8 & 2.46 $\pm$ 0.10 & 86.7 & 3.61 \\
211.0 & 9044-9054 & 9044-9047, 9054 & 5 & 0,4 & 26 & 2.18 $\pm$ 0.01 & 43.8 & 1.68 & 24 & 2.07 $\pm$ 0.03 & 144.6 $\pm$ 24.9 & 2.57 $\pm$ 0.13 & 12.0 & 0.50 \\
213.0 & 9070-9074 & 9070-9072, 9074 & 4 & 0,2,4,6 & 54 & 2.21 $\pm$ 0.02  & 55.7 & 1.03 & 52 & 2.10 $\pm$ 0.04 & 125.9 $\pm$ 39.1 & 2.41 $\pm$ 0.11 & 44.5 & 0.86 \\
218.0 & 9098-9110 & 9104-9105, 9107-9110 & 6 & 0,1 & 26 & 2.14 $\pm$ 0.01 & 44.0 & 1.69 & 24 &  2.13 $\pm$ 0.04 & 99.6 $\pm$ 143.8 & 2.17 $\pm$ 0.06 & 43.6 & 1.82 \\
220.0 & 9116-9119 & 9116-9119 & 4 & 1,2,3 & 40 & 2.08 $\pm$ 0.02 & 58.8 & 1.47 & 38 & 2.02 $\pm$ 0.08 & 75.8 $\pm$ 51.9 & 2.13 $\pm$ 0.05 & 56.9 & 1.50 \\
221.0 & 9121-9130 & 9121-9130 & 10 & 0,2,4,6 & 54 & 2.20 $\pm$ 0.01 & 115.1 & 2.13 & 52 & 2.13 $\pm$ 0.03 & 134.0 $\pm$ 35.6 & 2.35 $\pm$ 0.07 & 99.5 & 1.91 \\
222.0 & 9132-9137 &  9132-9135 & 4 & 0,1,2 & 40 & 2.16  $\pm$ 0.02 & 36.3 & 0.91 & 38 & 2.07 $\pm$ 0.05 & 116.1 $\pm$ 39.7 &  2.33 $\pm$ 0.10 & 28.9 & 0.76 \\
224.0 & 9142-9152 & 9142-9143, 9145-9146 & 4 & 1,2,3 & 40 &  2.19 $\pm$ 0.02 & 65.3 & 1.63 & 38 & 2.07 $\pm$ 0.04 & 137.4 $\pm$ 32.3 & 2.49 $\pm$ 0.13 & 47.0 & 1.24 \\
226.0 & 9158-9166 & 9158-9164, 9166 & 8 & 1,3,5,7 & 54 & 2.21 $\pm$ 0.01 & 144.9 & 2.68 & 52 & 2.12 $\pm$ 0.04 & 97.2 $\pm$ 27.6 & 2.32 $\pm$ 0.06 & 133.8 & 2.57 \\
227.0 & 9168-9180 & 9168-9180 & 13 & 2 & 12 & 2.17 $\pm$  0.01 & 40.3 & 3.36 & 10 & 2.12 $\pm$ 0.02 & 85.0 $\pm$ 19.2 & 2.26 $\pm$ 0.04 & 22.2 & 2.22 \\
228.0 & 9182-9194 & 9182-9184, 9188-9192 & 8 & 2,6 & 26 & 2.15 $\pm$ 0.01 & 71.3 & 2.74 & 24 & 2.07 $\pm$ 0.03 & 84.2 $\pm$ 19.7 & 2.25 $\pm$ 0.04  & 53.7 & 2.24 \\
229.5 & 9212-9215 & 9212-9215 & 4 & 1,3,5,7 & 54 & 2.17 $\pm$ 0.02 & 120.7 & 2.23 & 52 & 2.06 $\pm$ 0.05 & 119.8 $\pm$ 34.8 & 2.36 $\pm$ 0.10 & 110.5 & 2.13 \\
231.0 & 9203-9208 & 9203-9208 & 6 & 5,7 & 26 &  2.09 $\pm$ 0.02 & 28.9 & 1.11 & 24 & 1.94 $\pm$ 0.08 & 68.3 $\pm$ 19.8 &  2.17 $\pm$ 0.04 & 18.6 & 0.78 \\
232.5 & 9226-9236 & 9226-9234 & 9 & 1,5 & 26 &  2.25 $\pm$ 0.01 & 75.4 & 2.90 & 24 & 2.23 $\pm$ 0.02 & 139.4 $\pm$ 96.3 & 2.35 $\pm$ 0.11 & 72.8 & 3.04 \\
301.0 & 9217-9222 & 9217-9222 & 6 & 0,1,4 & 40 & 2.20 $\pm$ 0.01   & 62.1 & 1.55 & 38 & 2.11 $\pm$ 0.02 & 125.7 $\pm$ 23.3 & 2.43 $\pm$ 0.08 & 33.3 & 0.88 \\
302.3 & 9240-9250 & 9240-9244  & 5 & 1,3,5,7 & 54 &  2.20  $\pm$  0.02 & 123.0 & 2.28 & 52 &  2.14 $\pm$ 0.04 & 154.4 $\pm$ 49.6 & 2.40 $\pm$ 0.11 & 113.8 & 2.19 \\
303.2 & 9253-9260 & 9254-9260 & 7 & 6,7 & 26 & 2.13 $\pm$ 0.01 & 56.9 & 2.19 & 24 & 2.06 $\pm$ 0.03 & 110.5 $\pm$ 35.5 & 2.27 $\pm$ 0.08 & 47.5 & 1.98 \\
304.0 & 9280-9284 & 9280-9283 & 4 & 4,5,6,7 & 54 & 2.2 $\pm$ 0.02  & 44.3 & 0.82 & 52 & 2.17 $\pm$ 0.04 & 164.9 $\pm$ 68.0 & 2.48 $\pm$ 0.19 & 38.8 & 0.75 \\
305.0 & 9286-9292 & 9287-9292 & 6 & 4,5,6,7 & 54 & 2.15 $\pm$ 0.02 & 106.4 & 1.97 & 52 &  2.12 $\pm$  0.03 & 151.8 $\pm$ 166.2 & 2.22  $\pm$ 0.13 & 104.4 & 2.01 \\
308.6 & 9315-9321 & 9315-9321 & 7 & 6,7 & 26 & 2.20 $\pm$ 0.02  & 41.2 & 1.58 & 24 & 2.14 $\pm$ 0.03 & 140.5 $\pm$ 49.5 & 2.44 $\pm$ 0.17 & 33.7 & 1.40 \\
310.0 & 9323-9333 & 9323-9333 & 11 & 0,4,6 & 40 & 2.21 $\pm$  0.01 & 118.1 & 2.95 & 38 & 2.13 $\pm$ 0.02 & 145.0 $\pm$ 28.9 & 2.46 $\pm$ 0.10 & 93.5 & 2.46 \\
314.0 & 9356-9367 & 9356-9357, 9360-9362, 9364-9367 & 9 & 5,7 & 26 & 2.13 $\pm$ 0.01 & 76.5 & 2.94 & 24 & 2.11 $\pm$ 0.03 & 118.4 $\pm$ 104.5 & 2.18 $\pm$ 0.06 & 75.4 & 3.14 \\
315.0 & 9369-9374 & 9369-9374 & 6 & 5,7 & 26 &  2.21 $\pm$ 0.02 & 25.5 & 0.98 & 24 & 2.15 $\pm$ 0.03 & 144.1 $\pm$ 54.7 & 2.45 $\pm$ 0.18 & 18.0 & 0.75 \\
316.0 & 9376-9383 &  9376, 9379-9383 & 6 & 5,7 & 26 & 2.17 $\pm$ 0.01 & 51.6 & 1.99 & 24 & 2.13 $\pm$ 0.02 & 308.2 $\pm$ 77.2 & 3.29 $\pm$ 0.76 & 30.9 & 1.29 \\
319.0 & 9413-9418 & 9413-9418 & 6 & 0,4 & 26 & 2.20 $\pm$ 0.02 & 17.5 & 0.67 & 24 & 2.18 $\pm$ 0.04 & 104.0 $\pm$ 105.9 & 2.25 $\pm$ 0.09 & 16.8 & 0.70 \\
320.0 & 9420-9425 & 9420-9425 & 6 & 0,1 & 26 & 2.22 $\pm$ 0.02 & 35.5 & 1.37 & 24 & 2.16 $\pm$ 0.03 & 173.5 $\pm$ 56.1 & 2.75 $\pm$ 0.41 & 25.7 & 1.07 \\
321.1 & 9392-9397 & 9392-9395, 9397 & 5 & 0,2,4,6 & 54 & 2.17 $\pm$ 0.02 & 74.1 & 1.37 & 52 & 2.01 $\pm$ 0.05 & 147.3 $\pm$ 31.0 & 2.48  $\pm$ 0.13 & 54.8 & 1.05 \\
322.0 & 9448-9460 & 9448-9449, 9457-9460 & 6 & 0,2 & 26 & 2.06 $\pm$ 0.02  & 46.6 & 1.79 & 24 &  1.94 $\pm$ 0.13 & 63.8 $\pm$ 37.7 & 2.10 $\pm$ 0.05 & 43.9 & 1.83 \\
323.0 & 9434-9446 & 9434-9438, 9441-9444, 9446 & 10 & 1,3,7 & 40 & 2.20 $\pm$ 0.01 & 53.9 & 1.35 & 38 &  2.10 $\pm$  0.03 & 129.4 $\pm$ 28.8 & 2.42 $\pm$ 0.09 & 30.8 & 0.81 \\
325.0 & 9469-9481 & 9469-9475, 9477, 9479-9481 & 11 & 0,2,4 & 40 & 2.17 $\pm$ 0.01 & 169.7 & 4.24 & 38 & 2.07 $\pm$ 0.02 & 130.4 $\pm$ 15.7 & 2.55 $\pm$ 0.09 & 106.9 & 2.81 \\
326.0 & 9483-9488 & 9483-9488 & 6 & 0,1,2 & 40 & 2.13 $\pm$ 0.02  & 64.3 & 1.61 & 38 &  2.01 $\pm$ 0.06 & 80.0 $\pm$ 24.5 & 2.24 $\pm$ 0.06 & 55.5 & 1.46 \\
332.0 & 9522-9537 & 9522-9531, 9533-9537 & 15 & 1,3,7 & 40 & 2.23  $\pm$ 0.01 & 170.0 & 4.25 & 38 & 2.11 $\pm$ 0.02 & 133.4 $\pm$ 17.8 & 2.62 $\pm$ 0.10 & 111.0 & 2.92 \\
335.5 & 9559-9564 & 9559-9560, 9562-9564 & 5 & 0,1,2 & 40 & 2.17 $\pm$ 0.02 & 51.5 & 1.29 & 38 &  2.10  $\pm$ 0.11 & 65.3  $\pm$ 61.5 & 2.20 $\pm$ 0.06 & 50.3 & 1.32 \\
337.0 & 9574-9592 & 9574, 9576-9579, 9583, 9586-9591 & 12 & 0,2,4 & 40 &  2.20 $\pm$ 0.01 & 137.4 & 3.43 & 38 & 2.11 $\pm$ 0.02 & 139.6 $\pm$ 26.3 &  2.50 $\pm$ 0.10 & 106.4 & 2.80 \\
338.5 & 9596-9614 & 9596-9600, 9602, 9609, 9612-9614 & 19 & 0,4,5 & 40 & 2.18 $\pm$ 0.01 & 88.2 & 2.21 & 38 & 2.64 $\pm$ 0.26 & 38.9 $\pm$ 7.2 & 2.13 $\pm$ 0.02 & 74.6 & 1.96 \\
339.0 & 9616-9628 & 9616-9617, 9621-9627 & 9 & 3,7 & 26 & 2.13 $\pm$  0.02 & 57.1 & 2.20 & 24 & 2.01 $\pm$ 0.04 & 100.8 $\pm$ 19.2 &  2.36 $\pm$ 0.08 & 33.4 & 1.39 \\
\enddata


\tablecomments{Pertinent information for the 73 VP spectra covering the 
period from TJD 8407 to 9628 included in this report. It includes VP number, 
the range of days and "clean" days included in the VP, "source viewing" 
LADs, and the best-fit parameters for both a single power-law or a broken 
power-law representing the data.}


\end{deluxetable} 

\begin{deluxetable}{cccccccc}

\tablenum{2}
\tablewidth{0pt}
\tablehead{
\colhead{VP} & \colhead{TJD}   & \colhead{LADs}   & \colhead{$\nu$} &
\colhead{K}  & \colhead{$\alpha$} & \colhead{$\chi^2$} &\colhead{$\chi^2/\nu$}
}
\startdata
5.0 & 8451 & 1,3,5,7 & 54 & 17.3 $\pm$ 0.3 & 2.23 $\pm$ .06 & 40.1 & 0.74\\
5.0 & 8452 & 1,3,5,7 & 54 & 10.9 $\pm$ 0.1 & 2.11 $\pm$ .04 & 37.0 & 0.68\\
5.0 & 8453 & 1,3,5,7 & 54 & 14.0 $\pm$ 0.2 & 2.17 $\pm$ .04 & 42.6 & 0.79\\
5.0 & 8455 & 1,3,5,7 & 54 & 23.7 $\pm$ 0.3 & 2.29 $\pm$ .05 & 50.6 & 0.94\\
5.0 & 8456 & 1,3,5,7 & 54 & 16.0 $\pm$ 0.2 & 2.21 $\pm$ .05 & 35.5 & 0.66\\
5.0 & 8457 & 1,3,5,7 & 54 & 21.1 $\pm$ 0.3 & 2.27 $\pm$ .05 & 43.1 & 0.80\\
5.0 & 8458 & 1,3,5,7 & 54 & 16.6 $\pm$ 0.2 & 2.20 $\pm$ .05 & 62.2 & 1.15\\
5.0 & 8459 & 1,3,5,7 & 54 & 11.9 $\pm$ 0.2 & 2.14 $\pm$ .05 & 22.5 & 0.42\\
5.0 & 8460 & 1,3,5,7 & 54 & 14.1 $\pm$ 0.2 & 2.18 $\pm$ .05 & 35.3 & 0.65\\
5.0 & 8461 & 1,3,5,7 & 54 & 16.5 $\pm$ 0.2 & 2.21 $\pm$ .05 & 15.6 & 0.29\\
5.0 & 8462 & 1,3,5,7 & 54 & 18.6 $\pm$ 0.2 & 2.24 $\pm$ .05 & 38.0 & 0.70\\
5.0 & 8451-8453, 8455-8462 & 1,3,5,7 & 54 & 16.4 $\pm$ 0.9 & 2.21 $\pm$ .01 & 172.6 & 3.20\\
\enddata

\tablecomments{Best-fit power-law model parameters for each of the eleven single-day spectra in VP-5 are compared with those for the VP spectrum. While the power-law model (KE$^{-a}$) fits the single-day spectra well, because of the short-term spectral and flux variability, a simple power-law model cannot adequately fit the integrated VP spectrum.}

\end{deluxetable}


\begin{deluxetable}{ccccc}

\tabletypesize{\footnotesize}
\tablenum{3}
\tablecaption{Average gamma-ray fluxes for each of the three 400-day periods and their sum (photons cm$^{-2}$-s$^{-1}$-keV$^{-1}$)}
\tablewidth{0pt}
\tablehead{
\colhead{Energy (keV)} & \colhead{Period 1}   & \colhead{Period 2}   & \colhead{Period 3} & \colhead{
Periods 1-3}\\
\colhead{} & \colhead{TJD 8400-8800} & \colhead{TJD 8800-9200} & \colhead{TJD 9200-9628} & \colhead{TJD 8400-9628}\\
\colhead{} & \colhead{187-day Integration} & \colhead{197-day Integration} & \colhead{220-day Integration} & \colhead{604-day Integration}}
\startdata
35-40 & (543.37 $\pm$ 1.96) x 10$^{-5}$ & (511.81 $\pm$ 1.83) x 10$^{-5}$ & (528.65 $\pm$ 2.01) x 10$^{-5}$ & (527.19 $\pm$ 1.11) x 10$^{-5}$ \\
40-45 & (392.26 $\pm$ 1.56) x 10$^{-5}$ & (407.32 $\pm$ 1.55) x 10$^{-5}$ & (400.16 $\pm$ 1.59) x 10$^{-5}$ & (399.94 $\pm$ 0.91) x 10$^{-5}$ \\
45-55 & (309.91 $\pm$ 1.15) x 10$^{-5}$ & (293.47 $\pm$ 1.10) x 10$^{-5}$ & (305.48 $\pm$ 1.20) x 10$^{-5}$ & (302.63 $\pm$ 0.67) x 10$^{-5}$ \\
55-73 & (180.34 $\pm$ 0.68) x 10$^{-5}$ & (177.77 $\pm$ 0.64) x 10$^{-5}$ & (178.67 $\pm$ 0.71) x 10$^{-5}$ & (178.88 $\pm$ 0.39) x 10$^{-5}$ \\
73-98 & (98.74 $\pm$ 0.41) x 10$^{-5}$ & (97.91 $\pm$ 0.38) x 10$^{-5}$ & (98.21 $\pm$ 0.42) x 10$^{-5}$ & (98.27 $\pm$ 0.23) x 10$^{-5}$ \\
98-123 & (55.58 $\pm$ 0.27) x 10$^{-5}$ & (55.31 $\pm$ 0.26) x 10$^{-5}$ & (55.91 $\pm$ 0.29) x 10$^{-5}$ & (55.58 $\pm$ 0.16) x 10$^{-5}$ \\
123-162 & (31.74 $\pm$ 0.17) x 10$^{-5}$ & (31.11 $\pm$ 0.17) x 10$^{-5}$ & (31.59 $\pm$ 0.19) x 10$^{-5}$ & (31.47 $\pm$ 0.10) x 10$^{-5}$ \\
162-230 & (14.79 $\pm$ 0.10) x 10$^{-5}$ & (14.70 $\pm$ 0.10) x 10$^{-5}$ & (14.93 $\pm$ 0.12) x 10$^{-5}$ & (14.80 $\pm$ 0.06) x 10$^{-5}$ \\
230-313 & (69.55 $\pm$ 0.75) x 10$^{-6}$ & (70.14 $\pm$ 0.76) x 10$^{-6}$ & (68.75 $\pm$ 0.88) x 10$^{-6}$ & (69.55 $\pm$ 0.46) x 10$^{-6}$ \\
313-429 & (31.73 $\pm$ 0.56) x 10$^{-6}$ & (32.35 $\pm$ 0.57) x 10$^{-6}$ & (31.56 $\pm$ 0.66) x 10$^{-6}$ & (31.91 $\pm$ 0.34) x 10$^{-6}$ \\
429-595 & (16.57 $\pm$ 0.58) x 10$^{-6}$ & (16.89 $\pm$ 0.58) x 10$^{-6}$ & (16.71 $\pm$ 0.67) x 10$^{-6}$ & (16.72 $\pm$ 0.35) x 10$^{-6}$ \\
595-766 & (7.98 $\pm$ 0.40) x 10$^{-6}$ & (7.92 $\pm$ 0.40) x 10$^{-6}$ & (7.84 $\pm$ 0.45) x 10$^{-6}$ & (7.92 $\pm$ 0.24) x 10$^{-6}$ \\
766-1104& (4.60 $\pm$ 0.31) x 10$^{-6}$ & (4.66 $\pm$ 0.33) x 10$^{-6}$ & (4.66 $\pm$ 0.36) x 10$^{-6}$ & (4.64 $\pm$ 0.19) x 10$^{-6}$ \\
1104-1700 & (2.11 $\pm$ 0.23) x 10$^{-6}$ & (2.32 $\pm$ 0.23) x 10$^{-6}$ & (2.19 $\pm$ 0.26) x 10$^{-6}$ & (2.21 $\pm$ 0.14) x 10$^{-6}$ \\
\enddata

\end{deluxetable}

\end{center}

\begin{references}
\reference{Agrinier et al. 1990}Agrinier , B., et al. 1990, ApJ, 355, 645
\reference{Arnaud 1996}Arnaud, K. A., 1996, in ASP Conf. Proc. 101, Astronomical Data Analysis Software and Systems V, ed. G. Jacoby $\&$ J. Barnes (San Francisco; ASP), 17.
\reference{Ayre et al. 1983}Ayre, C. A., Bhat, P. N., Ma, Y. Q., Myers, R. M., and Thompson, M. G., 1983, MNRAS, 205, 285
\reference{Baker et al 1973} Baker, R. E., Lovett, R. R., Orford, K. J.,
 Ramsden, D., 1973, Nature, 245, 18.
\reference{Carpenter, Coe $\&$ Engel 1976}Carpenter, G. F., Coe, M. J., and Engel, A. R. 1976, Nature, 259, 99.
\reference{Dolan et al. 1977}Dolan, J. F., Crannel, C. J., Dennis, B. R., Frost, K. J., Maurer, G. S., and Orwig, L. E. 1977, ApJ, 217, 809
\reference{Fierro et al. 1998}Fierro, J. M., Michelson, P. F., and Nolan P. L., 1998, ApJ, 494, 734.
\reference{Fishman et al.1989}Fishman, G. J., et al. 1989, in Proc. Gamma Ray Observatory Science Workshop, ed. W.Johnson (Greenbelt: GSFC), 2.
\reference{Gilfanov et al. 1994}Gilfanov, M., et al., 1994, ApJS, 92, 411
\reference{Graser \& Schoenfelder 1982}Graser, U., and Schoenfelder, V. 1982, ApJ, 263, 677.
\reference{Gruber $\&$ Ling 1977}Gruber, D. E., \& Ling, J. C., 1977, ApJ, 213, 802.
\reference{Hameury et al. 1983}Hameury, J. M., Boclet, D., Durouchoux, Ph., Cline, T. L., Teegarden, B. J., Tueller, J., Paciesas, W. S., Haymes, R. C., 1983, ApJ, 270, 144.
\reference{Harmon et al. 2002}Harmon, B. A., Fishman, G. J., Wilson, C. A., Paciesas, W. S., Zhang, S. N., Finger, M. H., Koshut, T. M., McCollough, M. L., Robinson, C. R., and Rubin, B. C. 2002, ApJS, 138, 149.
\reference{Harnden \& Seward 1984}Harnden, F. R., Jr., and Seward, F. D. 1984, ApJ, 283, 279
\reference{Hasinger, 1984}Hasinger, G. 1984, MPE Rept. 186.
\reference{Hester et al. 1995} Hester, J. J., 1995, ApJ, 448, 240.
\reference{Hester et al. 2002} Hester, J. J., 2002, ApJ, 577, L49.
\reference{Hiller et al. 1970} Hillier, R. R., Jackson, W. R., Murray, A.,
 Redfern, R. M., Sale, R. G, 1970, ApJ., 162, 177.
\reference{Jung, 1989}Jung, G.V., 1989, ApJ, 338, 972.
\reference{Kennel \& Coroniti 1984}Kennel, C. F., \& Coroniti, F. V., 1984, ApJ, 365, 224
\reference{Knight 1982}Knight, F. K.1982, ApJ, 268, 538.
\reference{Kuiper et al. 2001}Kuiper, L., et al. 2001, A\&A, 378, 918.
\reference{Kurfess, 1971}Kurfess, J. D. 1971, ApJ (Letters), 168, L39.
\reference{Laros, Matteson and Pelling}Laros, J. G., Matteson, J. L., and Pelling, R. M., 1973, Nature Phys. Sci., 246, 109.
\reference{Leventhal, MacCallum $\&$ Watts, 1977}Leventhal, M. MacCallum, C. J., and Watts, A. C. 1977, Nature, 266, 6965.
\reference{Lin et al. 2003}Lin, R. et al., 2003, Solar Physics, in press.
\reference{Ling 2001}Ling, J. C., 2001, in AIP Conf. Proc. 280, Gamma-Ray Astrophysics 2001 Symp., ed. S. Ritz, N. Gehrels, \& C. R. Shrader (New York: AIP), 135
\reference{Ling et al. 1977}Ling, J. C., Mahoney, W. A., Willett, J. B., $\&$ Jacobson, A. S. 1977, Nature, 270, 36.
\reference{Ling et al. 1979 } Ling, J. C., Mahoney, W. A., Willett, J. B., $\&$ Jacobson, A. S., 1979, ApJ, 231. 896.
\reference{Ling et al. 1996}Ling, J. C., Wheaton, W. A., Mahoney, W. A., Skelton, R. T., Radocinski, R. G., and Wallyn, P. 1996, A\&AS, 120, 667.
\reference{Ling et al. 1997}Ling, J. C., et al.1997, ApJ, 484, 375
\reference{Ling et al. 2000}Ling, J. C., et al. 2000, ApJS, 127, 79
\reference{Ling \& Dermer. 1991}Ling, J. C. \& Dermer, C. D., 1991, in AIP Conf. Proc. 232, Gamma-Ray Line Astrophysics ed. Ph. Durouchoux \& N. Prantzos (Paris; Saclay), 361.
\reference{Mahoney, Ling $\&$ Jacobson. 1984}Mahoney, W. A., Ling, J. C., $\&$ Jacobson, A. S. 1984, ApJ, 278, 784.
\reference{Manchanda et al. 1982}Manchanda, r. K., Bazzano, A., La Padula, C. D., Polacaro, V. F., and Ubertini, P., 1982, ApJ, 252, 172
\reference{McConnell et al. 1987}McConnell, M. L., Dunphy, P. P.,Forrest, D. J., Chupp, E. L., Owens, A., 1987, ApJ, 321, 543.
\reference{McConnell et al. 2000}McConnell, M. L. et al., 2000, ApJ, 543, 928.
\reference{McConnell et al. 2002}McConnell, M. L. et al., 2002, ApJ, 572, 984.
\reference{Massaro et al. 1991}Masssaro, E., et al., 1991, ApJ, 376, L11.
\reference{Mori et al. 2003}Mori, K., Burrows, D. N., Hester, J. J., $\&$ Tsunemi, H., 2003, Bulletin of American Astronomical Society Meeting 201, 144.08.
\reference{Much et al. 1996}Much, R. P., et al., 1996, A\&AS 120, 703.
\reference{Nolan et al. 1993}Nolan et al., 1993, ApJ, 409, 697.
\reference{Owen. 1991}Owen. A., 1991, in AIP Conf. Proc. 232, Gamma-Ray Line Astrophysics, ed. Ph. Durouchoux \& N. Prantzos (Paris; Saclay), 341.
\reference{Parlier et al. 1991}Parlier, B., et al., 1991, in AIP Conf. Proc. 232, Gamma-Ray Line Astrophysics, ed. Ph. Durouchoux \& N. Prantzos (Paris; Saclay), 335.
\reference{Pelling et al. 1987}Pelling, R. M., Paciesas, W., Peterson, L. E., Markishima, K., Oda, M., Ogawara, Y., and Miyamoto, S., 1987, ApJ, 319, 416.
\reference{Philips et al. 1996}Philips, B. F., et al, 1996, ApJ, 465, 907.
\reference{Strickmann et al. 1979}Strickmann, M. S., Johnson,W. N., and Kurfess, J. D. 1979, Apl. J., 230, L15.
\reference{Strickman et al. 1982}Strickman, M. S., Kurfess, J. D., and Johnson, W. N.,1982, ApJ, 253, L23
\reference{Ubertini et al. 1974}Ubertini, P. et al. 1994, ApJ, 421, 269
\reference{Ulmer et al. 1995}Ulmer, M. P. et al., 1995, ApJ, 448, 356.
\reference{van den Bergh \& Pritchet 1989}van den Bergh, S., \& Pritchet, C. J., 1989, ApJ., 338, L69.
\reference{van der Meulen et al. 1998}van der Meulen, R. D., Bloemen, H., Bennett, K., et al., 1998, A\&A 330, 321. 
\reference{Wallyn et al. 2001} Wallyn, P., Ling, J. C., Mahoney, W. A.,
 Wheaton, W. A., Durouchoux, P., 2001, ApJ, 559, 342.
\reference{ Walraven et al. 1975}Walraven, G. D., Hall, R. D., Meegan, C. A., Colman, P. L., Shelton, D. H., and Haymes, R. C. 1975, ApJ, 202, 502.
\reference{Watanabe 1985}Watanabe, H., 1985, Ap. Space Sci., 111, 157
\reference{Weisskopt et al. 2000}Weisskopf, M. C., 2000, ApJ, 536, L81.
\reference{Willingale et al. 2001}Willingale, R., et al., 2001, A\&A, 365, L212.
\reference{Yoshimori et al. 1979}Yoshimori, M., Watanabe, H., Okudaira, K., Hirasima, Y., and Murakami, H., 1979, Aust. J. Phys., 32, 375.
\end{references}
\end{document}